\documentclass[acmsmall,review=false]{acmart}
\usepackage{hyperref}
\usepackage{latexsym}
\usepackage{mathrsfs}
\usepackage{multirow}
\usepackage{amsmath}
\usepackage{amsfonts}
\usepackage{graphicx}
\usepackage{tabularx}
\usepackage[boxed,ruled,lined]{algorithm2e}
\usepackage{algorithmic}
\usepackage{comment}
\usepackage{balance}
\usepackage{bm}

\usepackage{subcaption}
\usepackage{soul}

\newtheorem{theorem}{Theorem}

\newcommand{\xc}[1]{{\color{blue} xuchen: ``#1''}}

\AtBeginDocument{%
  \providecommand\BibTeX{{%
    \normalfont B\kern-0.5em{\scshape i\kern-0.25em b}\kern-0.8em\TeX}}}

\setcopyright{acmlicensed}
\acmJournal{TOIS}
\acmYear{2024} \acmVolume{1} \acmNumber{1} \acmArticle{1} \acmMonth{1}\acmDOI{XXXXXXX.XXXXXXX}

\begin{document}

\title{Regret-aware Re-ranking for Guaranteeing Two-sided Fairness and Accuracy in Recommender Systems}


\author{Xiaopeng Ye}
\affiliation{%
  \institution{Gaoling School of Artificial Intelligence}
   \country{Renmin University of China}
  \\	xpye@ruc.edu.cn
}

\author{Chen Xu}
\affiliation{%
  \institution{Gaoling School of Artificial Intelligence}
    \country{Renmin University of China}
    \\xc\_chen@ruc.edu.cn
}

\author{Jun Xu}
\authornote{Jun Xu is the corresponding author.}
\affiliation{%
  \institution{Gaoling School of Artificial Intelligence}
    \country{Renmin University of China}
  \\junxu@ruc.edu.cn
}

\author{Xuyang Xie}
\affiliation{%
   \country{Huawei Noah's Ark Lab}
  \\xiexuyang@huawei.com
}

\author{Gang Wang}
\affiliation{%
   \country{Huawei Noah's Ark Lab}
  \\wanggang110@huawei.com
}

\author{Zhenhua Dong}
\affiliation{%
   \country{Huawei Noah's Ark Lab}
  \\dongzhenhua@huawei.com
}

\begin{abstract}
In multi-stakeholder recommender systems (RS), users and providers operate as two crucial and interdependent roles, whose interests must be well-balanced.
Prior research, including our work BankFair, has demonstrated the importance of guaranteeing both provider fairness and user accuracy to meet their interests.
However, when they balance the two objectives, 
another critical factor emerges in RS: individual fairness, which manifests as a significant disparity in individual recommendation accuracy, with some users receiving high accuracy while others are left with notably low accuracy. This oversight severely harms the interests of users and exacerbates social polarization. How to guarantee individual fairness while ensuring user accuracy and provider fairness remains an unsolved problem.  
To bridge this gap, in this paper, we propose our method BankFair+. Specifically, BankFair+ extends BankFair with two steps: (1) introducing a non-linear function from regret theory to ensure individual fairness while enhancing user accuracy.
(2) formulating the re-ranking process as a regret-aware fuzzy programming problem to meet the interests of both individual user and provider, therefore balancing the trade-off between individual fairness and provider fairness.   Experiments on two real-world recommendation datasets demonstrate that BankFair+ outperforms all baselines regarding individual fairness, user accuracy, and provider fairness.
\end{abstract}


\maketitle
 
\section{Introduction}\label{sec:intro}
To create an equitable and sustainable multi-stakeholder recommendation system (RS),  the interests of both users and providers must be carefully balanced ~\cite{rochet2004two-sided_overview,wu2021tfrom,patro2020fairrec}. On the provider side, prior research~\cite{xu2024fairsync,patro2020fairrec,surer2018multistakeholder} has emphasized the importance of ensuring adequate item exposure within a period to maintain provider loyalty and engagement. On the user side, maintaining recommendation accuracy above a minimum threshold is crucial to sustain user engagement~\cite{ye2024bankfair,wu2021tfrom}.

\begin{figure}
    \centering
    \includegraphics[width=\linewidth]{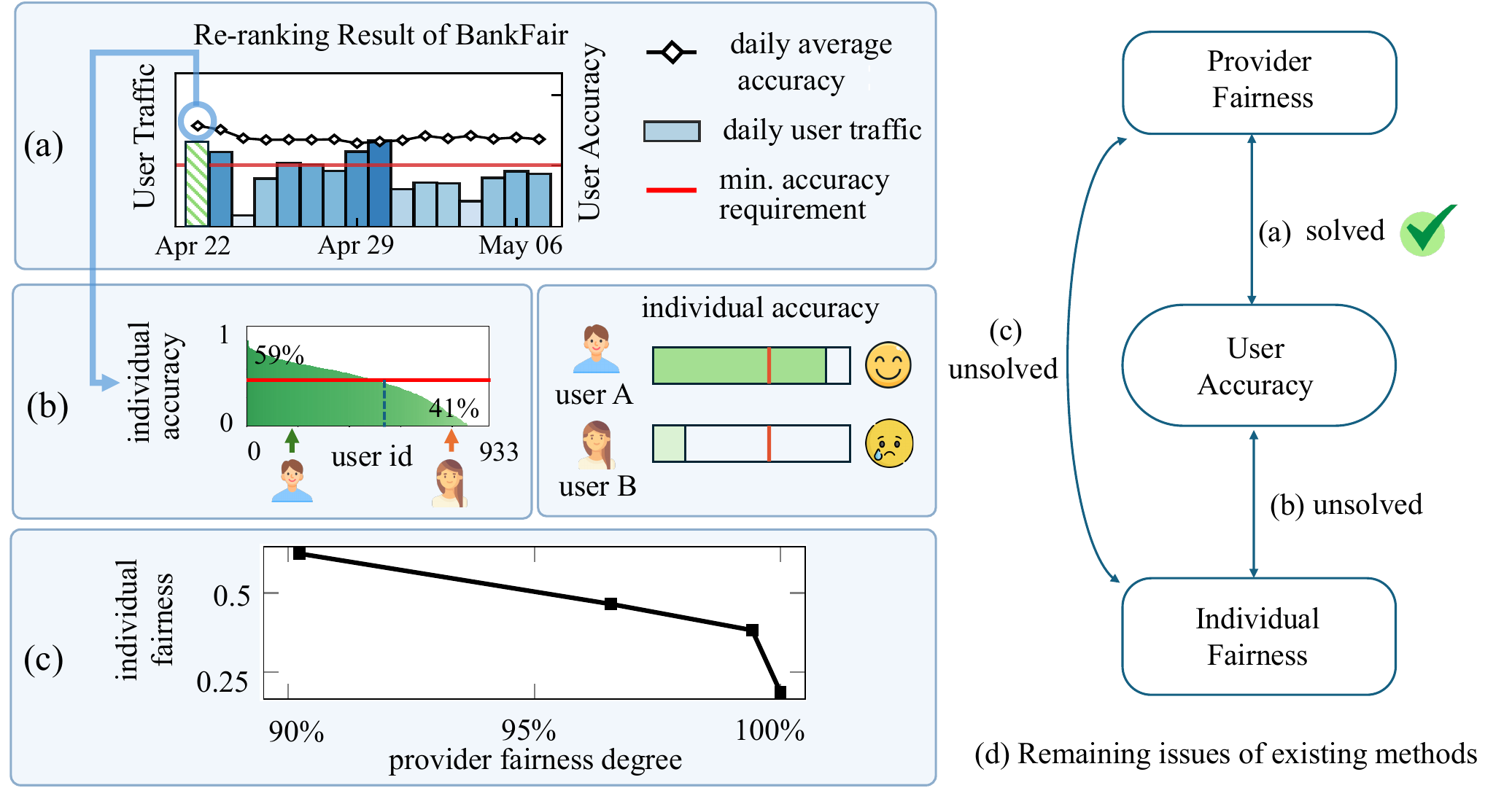}
    \caption{Although (a) previous works can guarantee both the user average accuracy and provider fairness, (b) severe individual unfairness of user accuracy still exists. (c) With the increase in provider fairness degree, such individual-level unfairness tends to escalate. (d) The remaining issues of existing methods: how to ensure individual fairness while guaranteeing provider fairness and user accuracy. }
    \label{fig:intro}
\end{figure}

Previous studies, including our previous work BankFair~\cite{ye2024bankfair}\footnote{Accepted by CIKM 2024}, have highlighted the necessity of ensuring provider fairness while maintaining user accuracy to satisfy their interests. BankFair primarily focuses on a real-world scenario where the user traffic (i.e., arriving user number within a period) inevitably fluctuates. Under this condition, 
BankFair revealed that user accuracy is more likely to be sacrificed to ensure provider fairness during periods of low user traffic. To address this challenge, BankFair employed the bankruptcy problem~\cite{curiel1987bankruptcygames} to allocate provider exposure dynamically across periods, which leverages surplus fairness during high-traffic periods to offset deficits during low-traffic periods, thereby ensuring fairness while enhancing accuracy in low-traffic scenarios. 
However, as evidenced by the experimental results in Figure~\ref{fig:intro}~(a), although BankFair well-balanced the trade-off between provider fairness and user accuracy,  another critical issue arises in RS: user individual fairness, which manifests as a significant accuracy disparity among individual users, i.e., some users receive recommendations with high accuracy, while some receive recommendations with extremely low accuracy.

To better illustrate this critical issue, we conduct an experiment using BankFair~\cite{ye2024bankfair} on the KuaiRand-1K dataset~\cite{gao2022kuairand}. As shown in Figure~\ref{fig:intro}(b), we select the day with the highest average user accuracy within a month (i.e., April 22, 2022) and plot the distribution of recommendation accuracy received by all individual users, sorted from highest to lowest. Despite the average accuracy exceeding the minimum accuracy requirement (i.e., red line), about 41\% of users receive accuracy below this threshold. By comparison between users in the 41\% group (e.g., user B) and those in the other group (e.g., user A), there a significant disparities in recommendation accuracy, highlighting significant individual unfairness. Furthermore, Figure~\ref{fig:intro}(c) shows that user individual fairness (y-axis) is more prone to compromise as the degree of provider fairness (x-axis) increases. Without loss of generality, this trade-off between individual fairness and provider fairness is a pervasive issue observed in most existing provider-fair methods~\cite{wu2021tfrom,naghiaei2022cpfair,xu2023p}.

In summary, we can summarize the issues in Figure~\ref{fig:intro} (a-c) in Figure~\ref{fig:intro} (d). While some prior studies~\cite{ye2024bankfair, xu2023p, wu2021tfrom} have attempted to guarantee both user accuracy and provider fairness (corresponding to the edge (a) in  Figure~\ref{fig:intro} (d)), they have largely overlooked the significance of individual fairness and its balance with user accuracy (edge (b) in  Figure~\ref{fig:intro} (d)) and provider fairness (edge (c) in  Figure~\ref{fig:intro} (d)). 
As emphasized by the well-established regret theory in economics~\cite{loomes1982regrettheory,ye2024bankfair, kahneman2013prospecttheory}, once a user receives a low-quality recommendation, the user can leave a lasting negative experience and dissatisfaction. Furthermore, as the quality of recommendations falls below expectations, user dissatisfaction grows non-linearly (i.e., the regret aversion effect~\cite{loomes1982regrettheory}). Consequently, such individual unfairness will largely harm the experience and interest of certain users who receive low accuracy, leading to user attrition~\cite{bardhan2022more} and social polarization~\cite{wang2024intersectional}. 
To bridge this gap and balance individual fairness with provider fairness and user accuracy, we propose our two-sided fair re-ranking model named BankFair+. Specifically, BankFair+ stikes two differences compared to the existing BankFair method:(1) to solve the problem of edge (b) in Figure~\ref{fig:intro} (d), we draw inspiration from the well-established regret theory~\cite{loomes1982regrettheory} and introduce a non-linear function to capture the relationship between user satisfaction and recommendation accuracy; (2) to solve the problem of edge (c) in Figure~\ref{fig:intro} (d), we propose a fuzzy programming-based online learning algorithm to balance the interests of individual user and providers. 
By combining two modules, we can achieve user-friendly and provider-fair re-ranking effectively and efficiently

Our main contributions can be summarized as follows:
\begin{itemize}
    \item We emphasize the significance of guaranteeing user accuracy, individual fairness, and provider fairness in two-sided platforms.
    \item We propose a two-sided re-ranking method named BankFair, which formulates the re-ranking process as a regret-aware fuzzy programming problem and utilizes an online learning algorithm to solve it.  
    \item The extensive experiments demonstrate that our approach outperforms existing baselines on two datasets in terms of provider fairness, user accuracy, and user individual fairness.
\end{itemize}

This paper is an extended version of our previous work~\cite{ye2024bankfair} published at the CIKM 2024.
We expand the BankFair framework proposed in the original work to BankFair+, with major
extensions: (1) We consider the issue of individual fairness in multi-stakeholder RS and aim to balance it with provider fairness and user accuracy. (2) We incorporate regret theory in the re-ranking stage and introduce a regret-aware non-linear function to measure user satisfaction, replacing the original linear function. The original BankFair exhibits individual unfairness in user accuracy after provider-fair re-ranking. To address this issue, BankFair+ employs a non-linear function to replace the original user accuracy measure. (3) We extend the original linear programming-based re-ranking method to fuzzy programming to balance the trade-off between individual user satisfaction and provider fairness. This strategy makes BankFair+ can ensure that user fairness is not excessively compromised under different degrees of provider fairness.
(4) We conduct more experiments and analysis on two real-world industrial datasets. All of the experiments demonstrate the effectiveness of
BankFair+ in guaranteeing user accuracy, individual fairness, and provider fairness.

\section{Related Work}
\subsection{Fairness-aware Re-ranking in Two-sided Platform}
Fairness in RS has drawn much attention in recent years, which can be divided into three types based on the target stakeholders: user fairness~\cite{li2021user,leonhardt2018userfairness}, provider fairness~\cite{xu2023p,qi2022profairrec,patro2020fairrec, xu2024taxation}, and two-sided fairness~\cite{patro2020fairrec,naghiaei2022cpfair}. 

For user fairness, we categorize the existing research into two main categories: group user fairness and individual user fairness. Most of the studies on group user-fairness~\cite{yao2017beyondparity_groupfair, edizel2020fairecsys} focus on grouping users based on their sensitive attributes. In cases where sensitive attributes are unavailable, users can be grouped according to their activity levels~\cite{fu2020fairness-aware-explain}. The fairness objective here is to ensure that statistical measures, such as error rates~\cite{edizel2020fairecsys} and average recommendation performance~\cite{fu2020fairness-aware-explain}, are balanced across different user groups. Group fairness can be achieved by incorporating fairness constraints into the recommendation system’s objective function~\cite{fu2020fairness-aware-explain, li2021user}. On the other hand, individual user fairness can be assessed by measuring disparities in recommendation quality and the diversity of explanations provided to different individual users~\cite{fu2020fairness-aware-explain}.

For provider fairness, we can mainly categorize the studies according to the ideologies and forms of fairness: (1) max-min fairness~\cite{xu2023p}, which aims to ensure the interests of the worst-off providers;
(2) equity of attention~\cite{biega2018equity,wu2021tfrom,naghiaei2022cpfair,morik2020controlling} which lets the exposure received by providers be proportional to their utility; (3) minimum exposure guarantee~\cite{patro2020fairrec,biswas2021fairrecplus,ben2023learning,lopes2024recommendations,xu2024fairsync,surer2018multistakeholder, yang2023vertical}, which tends to ensure that the exposure for providers over a period exceeds a minimum threshold. 
Within the research line of minimum exposure guarantee fairness, 
some heuristic methods~ \cite{patro2020fairrec,biswas2021fairrecplus,yang2023vertical} were proposed to ensure the minimum exposure in several recommendation lists and amortize the accuracy loss among each user (e.g., greedy round robin~\cite{patro2020fairrec,biswas2021fairrecplus}).
Also, some other works~\cite{lopes2024recommendations,surer2018multistakeholder,xu2024fairsync} formulated the re-ranking problem with exposure constraint as Integer-Programming (IP) and adopted online optimization methods (e.g., sub-gradient descent~\cite{duchi2011subgradient}) to solve it.
In most existing works~\cite{patro2020fairrec,biswas2021fairrecplus,yang2023vertical,surer2018multistakeholder}, the minimum exposure value cannot be directly specified but is controlled by a hyperparameter, which is hard to adjust and control. 
Thus, FairSync~\cite{xu2024fairsync} and BankFair~\cite{ye2024bankfair} considered a more industrially practical setting, which guarantees an arbitrarily specified minimum exposure, and we also consider this scenario in our paper.

Two-sided fairness in recommendation, which aims to primarily ensure fairness for both users and providers, has attracted much attention~\cite{wu2021tfrom, wang2024intersectional, naghiaei2022cpfair}. In particular, most studies emphasize user fairness based on performance (i.e., providing consistent recommendation quality to different user groups~\cite{wang2024intersectional} or individuals~\cite{wu2021tfrom}) and provider fairness related to exposure. Many of these studies~\cite{wu2021tfrom, wang2024intersectional, naghiaei2022cpfair} focus on fair re-ranking techniques, as allocating exposure is more practical during the re-ranking process, while~\citet{wu2022multi_obj} introduces a multi-objective optimization approach at the ranking stage. TFROM~\cite{wu2021tfrom} and CPFair~\cite{naghiaei2022cpfair} considered user individual and group fairness respectively in the context of provider-fair re-ranking, which both propose a Linear Programming (LP)-based method to ensure the two-sided fairness. Although many works have discussed the trade-off between user accuracy and provider fairness~\cite{ye2024bankfair, wu2021tfrom, naghiaei2022cpfair, xu2023p}. However, few works overlooked the trade-off between provider fairness and user fairness. In this paper, we aims to discuss the trade-off between user accuracy, user individual fairness and provider fairness, and we try to balance them simultaneously.

\subsection{Regret Theory in Decision Making}
To mitigate individual unfairness while guaranteeing user accuracy, we introduce the non-linear function from regret theory~\cite{loomes1982regrettheory} to reflect the relationship between recommendation quality and user satisfaction.
Most existing fair re-ranking methods are based on the expected utility theory (i.e., trying to maximize the overall utility of users) and assume that user satisfaction is linear to the recommendation accuracy~\cite{ye2024bankfair}, which assumes that users are risk-neutral~\cite{ge2020userriskpref}. 
However, in real-world decision-making, individuals are often influenced by subjective preferences shaped by psychological and behavioral factors. One important framework that captures this aspect is regret theory, first introduced by \citet{bell1982regret} and \citet{loomes1982regret}. In recent years, there has been a growing interest in the research and application of regret theory\cite{somasundaram2017regret_attitude}. This theory posits that decision-makers not only evaluate the outcomes of their chosen options but also consider what the outcomes of alternative choices might have been. If they perceive that another option could have resulted in a better outcome, they experience psychological regret. Conversely, when their choice turns out to be the best option, they feel satisfaction. As a result, the perceived utility in decision-making is composed of two elements: the utility of the chosen outcome and the “regret-rejoice” value, which reflects a comparison with the potential outcomes of alternatives. Researchers like \citet{chorus2012regret_route} and \citet{qu2019algorithms} have pointed out that regret theory offers distinct advantages over cumulative prospect theory in decision-making. Unlike cumulative prospect theory~\cite{kahneman2013prospecttheory}, regret theory does not require predefined reference points and involves fewer parameters, making it more straightforward to compute~\cite{zhang2014risky_attribute}. In decision models, regret theory presents a more realistic approach compared to expected utility theory, aligning better with how humans make decisions~\cite{wang2020projection_regret}.

\subsection{Fuzzy Programming for Balancing Conflicting Objectives}
To guarantee both individual user accuracy and provider fairness, we introduce fuzzy programming ~\cite{zimmermann1978fuzzy} to replace the traditional linear programming (LP) approach~\cite{kantorovich1960mathematical_methods}, balancing the trade-off between the two objectives. 

In recent fair re-ranking works that related to us, most works utilize LP-based methods to conduct such re-ranking processes and make recommendations. For example, P-MMF~\cite{xu2023p} innovatively proposed an online linear programming algorithm to fairly allocate exposure resources to help the worst-exposed provider. BankFair~\cite{ye2024bankfair} also presented an online learning algorithm to guarantee the minimum exposure for each provider.

However, in two-sided RS, the traditional LP approach usually cannot balance individual user accuracy and provider fairness, leading to potential individual unfairness.  This is because when considering individual accuracy, we have to impose numerous constraints to ensure the accuracy of each user. The constraint number of the LP problem will be largely increased since the user number in RS is very large~\cite{xu2023p}, which will incur significant computational overhead. Moreover, since the constraints in LP must be clear~\cite{dantzig2002linearpro} rather than vague,  an LP problem with such excessive accuracy and fairness constraints often conflicts with each other and cannot obtain feasible solutions.

To overcome such weakness in dealing with an excessive number of constraints which might conflict with each other, FP offers a excellent solution. FP is a combination of fuzzy mathematics ~\cite{zadeh1965fuzzyset} and optimization techniques~\cite{dantzig2002linearpro}, designed to handle problems with uncertainty in objectives or constraints. It represents conflicting objectives as fuzzy sets~\cite{zimmermann1978fuzzy, zadeh1965fuzzyset} and quantifies their satisfaction using membership functions. This approach allows constraints to have a certain degree of flexibility, avoiding rigid restrictions. By applying fuzzy set theory~\cite{zadeh1965fuzzyset}, multiple objectives can be integrated into a comprehensive objective function or balanced using fuzzy weights.
Thanks to its advantages, FP theory~\cite{zadeh1965fuzzyset} has become a widely used tool in decision-making problems.  

\section{Preliminaries}\label{sec:formulation}


\subsection{Guranteeing Accuracy and Fairness in Recommender System}

\subsubsection{Recommender System with users and providers}
Let $\mathcal{U}, \mathcal{I}, \mathcal{P}$ denote the set of users, items, and providers, respectively. For a provider $p$, there are multiple items in the set $\mathcal{I}_p$ belonging to $p$. 
For each user-item pair $(u, i)$, RS generates a preference score $s_{u, i} \in [0,1]$ estimated from the user-item interaction history, which represents the degree of the user's preference for the item.
When a user $u \in \mathcal{U}$ visits RS, RS will first generate a ranking list $\pi^*(u,K)\in \mathcal{I}^K$ of size $K$ to maximize the preference score $s_{u, i}$ that users can receive.  
After the ranking phase, the goal of two-sided re-ranking is to compute a new list $\pi(u,K)$ which well balances the user accuracy, individual fairness, and provider fairness. 
We denote the quality of the re-ranking list $\pi(u,K)$ that user $u$ receives as $q_{\pi_u}$, which is usually measured by user accuracy based on the preference score that the user receives (e.g. DCG~\cite{jarvelin2002ndcg}).
For simplicity, we will represent $\pi(u,K)$ as $\pi_u$ in the following text.

 When an item $i$ is recommended in $\pi(u,K)$, then the corresponding provider $p$ with such an item (i.e., $i \in \mathcal{I}_p$) can have an exposure value of $p(\sigma(i))$, where $\sigma(i)$ is the rank of item $i$ in ranking $\sigma$, and $p(\cdot)$ casts the rank to the user's examine probability. 
The total exposure of provider $p$ after the recommendation of $\mathcal{U}$ will be:

\begin{equation}
\label{eq:exposure_with_decay}
     \bm{e}_p = \sum_{u\in\mathcal{U}}\sum_{i \in \pi(u,K)}p(\sigma(i)) \cdot \mathbb{I}(i\in \mathcal{I}_p),
\end{equation}  
 where $\mathbb{I}(\cdot)$ is the indicator function, which means $\mathbb{I}(\cdot)=1$ if condition $(\cdot)$ is met, otherwise 0. We use function $G(\bm{e})$ to denote the unfairness of provider exposure.

\subsubsection{Two-sided Re-ranking with User and Provider Constraint}
For a sequence of arriving users $\mathcal{U} =\{u_1, u_2, \cdots, u_n\}$, previous works~\cite{ye2024bankfair} have demonstrated the necessity of guaranteeing both accuracy and provider fairness in RS. ~\citet{ye2024bankfair} mainly focuses on ensuring average user accuracy and provider fairness under fluctuating user traffic. However, this approach may still result in some users receiving very low utility, reducing their experience and satisfaction.
Hence, in this paper, our goal is to propose an approach that can simultaneously ensure provider fairness and individual user accuracy.
Formally, we formulate this two-sided re-ranking task as a constrained optimization problem:
\begin{align}
\label{eq:unfied_opt}
f (\phi, \eta) =\quad &\max_{\pi} \frac{1}{|\mathcal{U}|} \sum_{u \in\mathcal{U}} Z(\pi_u) \quad  \quad &\text{(User Satisfaction Objective)}\\ \label{eq:unfied_opt_con1}
        \textrm{s.t.} \quad 
        &  q_{\pi_u}\ge \phi,\quad \forall u \in \mathcal{U}  \quad \quad &\text{(Individual User Accuracy Guarantee)}\\
\label{eq:unfied_opt_con3}
        & G(\bm{e}) \le \eta, \quad\quad &\text{(Provider Fairness Guarantee)}
    \end{align}
where $Z(\cdot)$ is the user satisfaction function which corresponds with the quality $q_{\pi_u}$ of the ranking list $\pi_u$, $G(\cdot)$ is the function to measure the unfairness of provider exposure $\bm{e}$. 
Next, we will introduce the three key parts of this task separately.


\begin{itemize}
    \item \textbf{ User Satisfaction Objective.} 
Objetive~\eqref{eq:unfied_opt} aims to maximize the average user satisfaction $Z(\bm{q}_u)$ of all users $u\in \mathcal{U}$. Previous re-ranking methods~\cite{ye2024bankfair, xu2023p, naghiaei2022cpfair} assume that the user satisfaction has a linear relationship with quality, i.e., $Z(\pi_u) = q_{\pi_u}, u\in \mathcal{U}$.

\item \textbf{Individual User Accuracy Guarantee.} 
Constraint~\eqref{eq:unfied_opt_con1} requires the accuracy $q_{\pi_u}$ of each user $u$ should be no less than the minimum accuracy requirement $\phi$. It is easy to see that when the accuracy $q_{\pi_u}$ of all individual users is ensured (i.e., $q_{\pi_u}\ge \phi$), the average user accuracy will also be guaranteed ($\frac{1}{|\mathcal{U}|} \sum_{u \in \mathcal{U}}q_{\pi_u}\ge \phi$).

\item \textbf{Provider Fairness Guarantee.} 
Constraint~\eqref{eq:unfied_opt_con3} demands the unfairness $G(\bm{e})$ of provider $p$ should not be more than the fairness requirement $\eta$.
\end{itemize}




\subsubsection{Re-ranking with Algorithm Updates under Fluctuating User Traffic.}
For the re-ranking problem $f(\phi, \eta)$, BankFair~\cite{ye2024bankfair} considered a real-world industrial-practical circumstance, where the user traffic (i.e., arriving user number within each algorithm update interval) is fluctuating and the RS needs to guarantee each provider a minimum exposure at each algorithm update interval. Next, we will introduce these two settings separately.

In real-world scenarios, different users  $\mathcal{U}_n=\{u_t| t_n\leq t <t_{n+1}, n\in [1,2,\ldots,N]\}$ will arrive in RS within time interval $n$, where $t_n$ is the start time of interval $n$. We use $\bm{r}_n=|\mathcal{U}_n|$ to denote the user traffic (i.e., user number) arriving within time interval $n$ (e.g. $n$ means $n$-th hour or $n$-th day during the recommendation process). 


Under such user traffic fluctuating conditions, RS usually demands the cumulative exposure $\bm{e}_p = \sum_{n=1}^N \bm{E}_{p,n}$ of provider $p$ within $N$ intervals should be no less than the required minimum exposure $\bm{m}_p$ (i.e., $\bm{e}\geq \bm{m}_p$), where $\bm{E}_{p,n}$ denotes the earned exposure for provider $p$ at interval $n$. 
Therefore, the provider fairness function $G(\bm{e})$ of problem $f(\phi, \eta)$ becomes the form of minimum exposure guarantee fairness, i.e.,  i.e., $G(\bm{e})  = \sum_{p} \max \{\bm{m}_{p} - \bm{e}_p, 0 \}$.
However, under real industrial settings, re-ranking model $f(\phi, \eta)$ updates its parameters regularly~\cite{zanardi2011dynamicupdating,wei2011differentupdateinterval},  which makes the ideal re-ranking problem $f(\phi, \eta)$ cannot be globally optimized for all intervals. 

Hence, decomposed sub-problems within each interval will be solved. BankFair proposes a two-module re-ranking strategy. In module 1, BankFair proposed a traffic-adaptive exposure allocation strategy to allocate the overall required minimum exposure $\bm{m}_p$ to each interval $n$ into $\bm{M}_{p,n}$ (i.e., $\bm{M}_{p,n}$ is the required minimum exposure for provider $p$ at interval $n$, and $\sum_{n=1}^N\bm{M}_{p,n}\ge\bm{m}_p$). In module 2, we solved the sub-problem $f(\phi, \eta)$ with constraint $\bm{E}_{p,n} \geq\bm{M}_{p,n}$ online in module 2.
Although BankFair conducted theoretical analysis and maximized the averaged user accuracy objective~\eqref{eq:unfied_opt} (The detail method of the module 1 of BankFair is in Section~\ref{sec:module1}), it has overlooked the individual unfairness issue (i.e., the individual user accuracy guarantee~\eqref{eq:unfied_opt_con1}) which the online learning recommendation algorithm causes, which will be taken into consideration in BankFair+. 




\subsection{Regret Theory}
\label{sec:regret_theory}
In this section, we will first introduce the regret theory. Then we will correspond the theory into the fair re-ranking task in recommendation.


In actual decision-making process, most of the decision-makers are not so rational, so the decision-maker's behavior factors need to be considered when making a decision. The prospect theory and regret theory are put forward in this context. In the increasingly complex, modern, political, and economic environment, decision-makers need to consider not only the results obtained after choosing a certain scheme but also the possible decision results after assuming that other alternatives are chosen. In the regret theory, the perceived utility function is composed of two parts: the utility function of current decision-making results and the regret-rejoice function compared with other decision-making results.

Let $\pi$ and $\pi'$, respectively, represent the results that can be obtained by selecting scheme $\pi$ and reference scheme $\pi'$. Then, the perceived utility of decision-makers for scheme $\pi$ is:

\begin{equation}
Z(\pi)=V(\pi)+R(V(\pi)-V(\pi')) .
\end{equation}

Among them, $V(\pi)$ represents the utility value of scheme $\pi$, and $R(V(\pi)-V(\pi'))$ is called regret-rejoice value. 

\begin{itemize}
    \item \textbf{Utility Function}. According to Loomes and Sugden~\cite{loomes1982regret}, the power function $V(\pi) = \pi^\alpha, 0 < \alpha < 1$ is used as a utility function of the attribute value in this paper. The greater the degree of risk aversion of decision-makers is, the smaller $\alpha$ is. $\alpha$ is called the risk aversion coefficient of decision-makers.

    \item  \textbf{Regret-rejoice Function}. This function measures the regret or rejoicing value of the decision-maker based on comparative psychology after making a choice.
If $R(V(\pi)-V(\pi'))$ is positive, then it is called a rejoice value, which indicates the extent to which the decision maker is glad to choose the scheme or give up the scheme. If $R(V(\pi)-V(\pi')) $ is negative, then it is called a regret value, which indicates the extent to which the decision maker regrets choosing the scheme or giving up the scheme. Obviously, the regret gratification function R(·) should be monotonically increasing and concave, i.e., it satisfies $R'(\cdot) > 0$, $R''(\cdot) < 0$, and $R(0) = 0$. 

Loomes and Sugden~\cite{loomes1982regret} pointed out that regret-rejoice function $R(\cdot)$ can be expressed as follows:

\begin{equation}
R(\Delta v)=1-\exp (-\delta \Delta V)
\end{equation}

Here, $\delta > 0$ is the regret avoidance coefficient of the decision maker, and the greater $\delta$ related to, the larger the regret avoidance degree of the decision maker. $\Delta v$ is the difference between the utility value of any two schemes. 
\end{itemize}

\subsection{Fuzzy Programming}


Fuzzy programming (FP) was first proposed by Zimmermann~\cite{zimmermann1978fuzzy} to solve multiple objective problems in which some objectives are conflicting, imprecise, or fuzzy in nature. FP has been applied successfully in many real-world problems, such as supply chain management, and engineering design, demonstrating particular effectiveness in scenarios requiring human preference incorporation and gradual constraint satisfaction. One advantage of FP is that the different shapes of membership functions can be used to describe different fuzzy goals, ensuring that each fuzzy goal can be well-balanced and not be excessively compromised. A FP problem with a solution set $x$, and fuzzy goals $z_k(x)$ and $g_s(\mathbf{x})$ related to $x$ is given by:
\begin{equation}
\begin{aligned}
&\begin{array}{ll}
z_k(\mathbf{x}) \gtrsim z_k^*, & \text { for } k=1,2, \ldots, l \\
g_s(\mathbf{x}) \lesssim g_s^0, & \text { for } s=1,2, \ldots, r
\end{array}\\
&\text { s.t. } \quad \mathbf{x} \in F \quad \text { (where } F \text { is a feasible set) }
\end{aligned}
\label{eq:fGP}
\end{equation}

where $z_k (x) \gtrsim z_k^*$ indicates that the i-th fuzzy goal 
is approximately greater than or equal to the aspiration level $z^*_k$.
$g_s(x)\lesssim g_s^0$  means that the i-th fuzzy goal is smaller than or equal to the aspiration level $g^0_s$.

In the compromised approach~\cite{chung2018fuzzy}, the FP problem~\eqref{eq:fGP} with $n$ objectives can be expressed as the following mathematical programming:

\begin{equation}
\operatorname{Min} \sum_{k=1}^l \lambda_k\left[\frac{z_k^*-z_k(x)}{z_k^*-z_k^{-}}\right]+\sum_{s=1}^\gamma \lambda_s\left[\frac{g_s(x)-g_s^*}{g_s^{-}-g_s^*}\right]
\end{equation}

where $\lambda_k>0$, $\lambda_s>0$ ( $\sum_{k=1}^l \lambda_k+\sum_{s=1}^r \lambda_s=1$) represent the weights of each objective function.

\section{Our Approach: BankFair+}
In this section, we present our re-ranking algorithm BankFair+ to guarantee user accuracy, and two-sided fairness. 
To balance provider fairness and user accuracy (i.e., the issue (a) in Figure~\ref{fig:intro} (d)), we introduce Module 1 (Section~\ref{sec:module1}). Then, we propose Module 2 with two steps. To balance individual fairness and user accuracy (i.e., the issue (b) in Figure~\ref{fig:intro} (d)), we introduce step 1 of Module 2 (Section~\ref{sec:step1}).
To balance individual fairness with provider fairness (i.e., the issue (c) in Figure~\ref{fig:intro} (d)), we introduce step 2 of Module 2 (Section~\ref{sec:step2}).

\begin{figure}
    \centering
\includegraphics[width=1\linewidth]{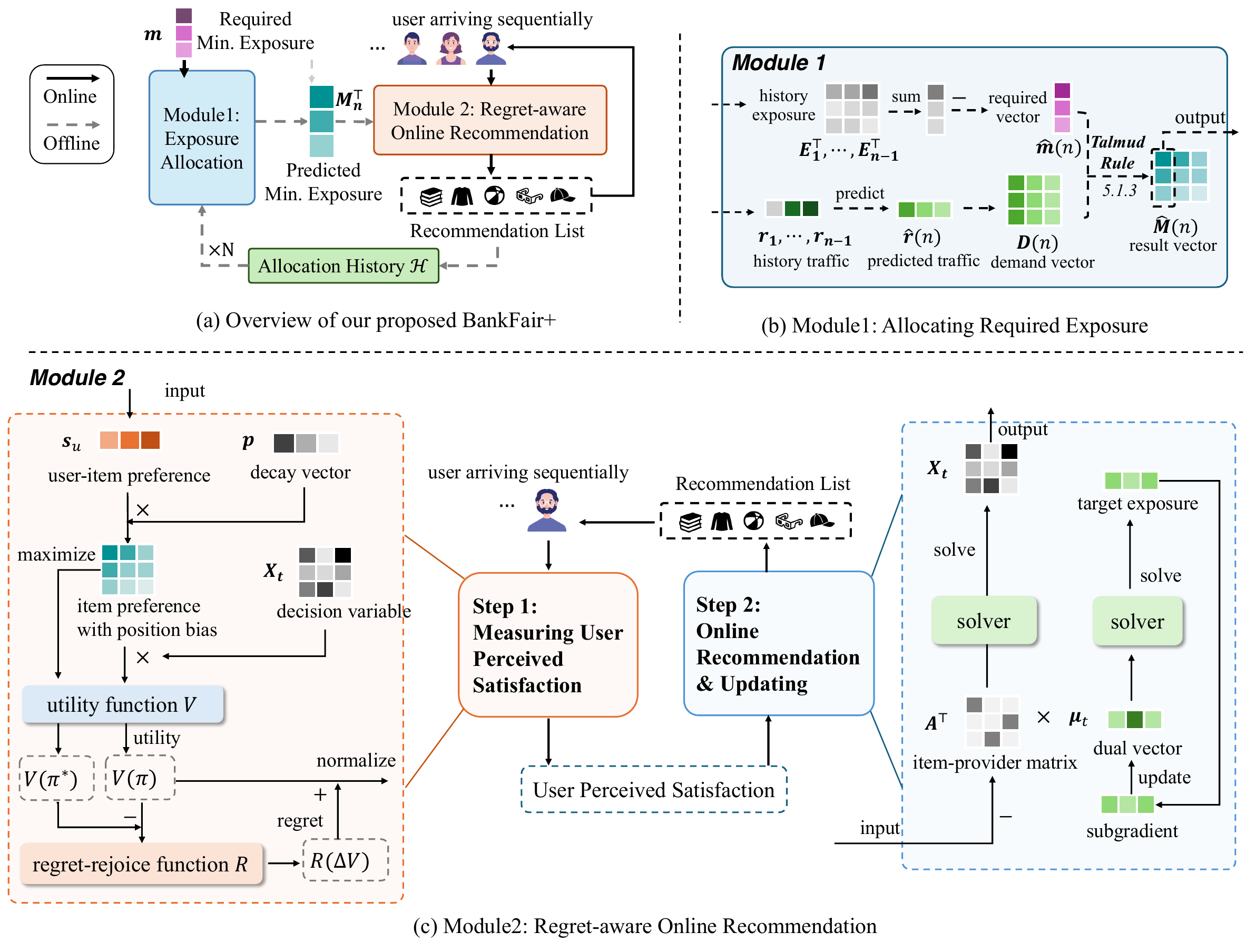}
    \caption{(1) Workflow of the proposed BankFair+ re-ranking algorithm. (b) In module 1, we allocate exposure to providers using the Talmud rule in bankruptcy problem. (c) In module 2, we propose a regret-aware online re-ranking algorithm to mitigate individual unfairness.}
    \label{fig:big}
\end{figure}

Figure~\ref{fig:big} depicts the overall workflow of BankFair+, and Algorithm~\ref{alg:bankfair} describes the overall procedure. The algorithm consists of the following two modules:
(1) \textit{Line 2-8}: module 1, which is identical to module 1 of BankFair~\cite{ye2024bankfair}, utilizes the bankruptcy problem to allocate minimum exposure to each interval, aiming to guarantee user accuracy and provider fairness at each inverval;
(2) \textit{Line 9-23}: module 2 proposes a two-step online learning algorithm. 
In step 1 (\textit{Line 13-18}), to ensure both individual fairness and user accuracy, we take inspiration from regret theory and introduce a non-linear function to measure user satisfaction. In step 2 (\textit{Line 19-23}), we utilize fuzzy programming to balance individual fairness and provider fairness goals.

\subsection{Module 1: Guaranteeing User Accuracy and Provider Fairness via Bankruptcy Problem}
\label{sec:module1}
In this section, we transform the traditional bankruptcy problem into a sequential problem to get the predicted minimum exposure $\bm{M}_n^{\top}$ at each interval $n$.

When the user traffic is low, adopting a larger minimum exposure will result in a greater accuracy loss~\cite{ye2024bankfair}. Therefore, to guarantee the accuracy level, we need to decrease the fairness degree (i.e., give a low demand for providers) at lower user traffic levels to reduce the accuracy loss. Otherwise, we need to increase the fairness degree (i.e., give a high demand for providers) to compensate for the low user traffic period, fulfilling the exposure requirements $\bm{m}$.

In the sequential bankruptcy problem of fair re-ranking, we update the input triplet (upcoming time intervals $\mathcal{N}=\{n,\cdots, N\}$, required minimum exposure $\hat{\bm{m}}_p(n)$, demanded minimum exposure $\bm{D}_p(n)$) and get the result vector $\hat{\bm{M}}(n)$ at each interval $n$.

In this module, we first predict $\bm{D}_p(n)$ in each interval $n$ to be the demand vector using the historical data $\mathcal{H}_{n-1}:=\{\bm{r}_{s},\bm{E}^{\top}_{s}\}_{s=1}^{n-1}$. Then we will use historical data $\mathcal{H}_{n-1}$ to update the remaining required minimum exposure $\hat{\bm{m}}_p(n)$ of the bankruptcy problem.
Finally, we use the input triplet to get the result vector $\hat{\bm{M}}(n)$ and output its first column as the predicted minimum exposure $\bm{M}_n^{\top}$.

\subsubsection{Update Remaining Minimum Exposure}
Then, we need to update the remaining required minimum exposure $\hat{\bm{m}}_p(n)$ at each interval $n$ as the estate for sequential bankruptcy problem. 
Specifically, at the beginning of each interval $n$, we collect the allocation history $\mathcal{H}_{n-1}:=\{\bm{r}_{s},\bm{E}^{\top}_{s}\}_{s=1}^{n-1}$ to calculate the remaining exposures $\hat{\bm{m}}_{p}(n)\in\mathbb{R}^+$ at interval $n$:
\begin{equation}
    \hat{\bm{m}}_{p}(n) =\left[\hat{\bm{m}}_{p}(n-1) - \bm{E}_{p,n-1} \right]_{+}=\left[\bm{m}_{p} -  \sum_{i=1}^{n-1}\bm{E}_{p,i} \right]_{+}, \forall p\in\mathcal{P},
\end{equation}
where $[\cdot]_{+} \triangleq \max \{0, \cdot\}$.

\subsubsection{Obtain Output Vector}

After getting the input triplet, we will utilize the Talmud rule~\cite{thomson2003axiomatic} in bankruptcy problem to obtain the output vector $\bm{M}_{n}^{\top}$ by getting the first column (i.e., present interval) of the matrix $\hat{\bm{M}}(n)$, i.e.,
$\bm{M}_{n}^{\top}  =  \hat{\bm{M}}_1(n)^{\top}$.
$\hat{\bm{M}}(n)\in\mathbb{R}^{|\mathcal{P}|\times |\mathcal{N}|}$ can be viewed as the predicted future minimum exposure,
which can be calculated by the Talmud rule:
\begin{equation}
\begin{aligned}
\hat{\bm{M}}_{p,i}(n) &= \text{TAL}_i\left(\mathcal{N},\hat{\bm{m}}_p(n),\bm{D}_p(n)\right)\\ & = \begin{cases}
\min\{\frac{\bm{D}_{p,i}(n)}{2},\theta\} & \text{if}~ \hat{\bm{m}}_p(n) \leq \frac{\bm{1}^{\top}\bm{D}_p(n)}{2} \\
\max\{\frac{\bm{D}_{p,i}(n)}{2},\bm{D}_{p,i}(n)-\theta\} & \text{if}~\hat{\bm{m}}_p(n) > \frac{\bm{1}^{\top}\bm{D}_p(n)}{2}
\end{cases},
\end{aligned} 
\label{eq:talmud}
\end{equation}
where $\theta$ is the parameter 
that makes the result vector $\hat{\bm{M}}_{p}(n)$ satisfying 
$\sum_{i\in \mathcal{N}} \hat{\bm{M}}_{p,i}(n)=\hat{\bm{m}}_p(n)$.

Through the Talmud rule $\text{TAL}(\cdot)$, we can allocate the required minimum exposure $\hat{\bm{m}}_p(n)$ into each interval according to the bankruptcy problem. 
The Talmud rule can be seen from two conditions: (1) If the minimum exposure $\hat{\bm{m}}_p(n)$ does not exceed half of the total demands $\bm{1}^{\top}\bm{D}_p(n)/2$: $\hat{\bm{m}}_p(n)$ will be distributed equally among each time interval $n$ until $\hat{\bm{M}}_{p,n}(n)$ meets half of the demand $\bm{D}_{p,n}(n)/2$.
(2) If the minimum exposure $\hat{\bm{m}}_p(n)$ exceed the total demands $\bm{1}^{\top}\bm{D}_p(n)/2$: $\hat{\bm{m}}_p(n)$ will be allocated to equalize the difference between each allocation $\hat{\bm{M}}_{p,n}(n)$ and its demand $\bm{D}_{p,n}(n)$.

\subsection{Module 2: Regret-aware Online Recommendation}
After getting the predicted minimum exposure at each interval $n$ in Module 1 (see Section~\ref{sec:module1}), we utilize the it as the required exposure for each provider at each interval $n$ and combine it with user accuracy goal to balance the user accuracy and two-sided fairness. Module 2 consists of two steps. In step 1, we integrate the well-known regret theory into the re-ranking process and propose a non-linear function to mitigate inidividual unfairness and guarantee user accuracy. In step 2, we formulate the re-ranking process as fuzzy programming~\cite{zimmermann1978fuzzy} to combine the goal of user individual accuracy and provider fairness.
Finally, we adapt the offline fuzzy algorithm into an online version to adapt the sequential arriving characteristic of users.

\subsubsection{\textbf{Step 1: Guaranteeing Individual Fairness and User Accuracy via Regret Theory}}
\label{sec:step1}
To ensure individual fairness while enhancing user accuracy, we implement regret theory in re-ranking scenarios. We will next map the three key functions (i.e., perceived utility function, utility function, and regret-rejoice function) in regret theory into the re-ranking process of RS.

To introduce the regret theory into the RS, we view the recommendation re-ranking process as a decision-making process, with the platform acting as the decision-maker. Each time the platform performs a fairness-aware re-ranking for a user $u$, it can be considered as a decision made by the platform. To meet fairness objectives $G(\bm{e})$, the platform may sacrifice some recommendation quality $q_{\pi_u}$, which will affect each user’s recommendation list $\pi_u$. After receiving the recommendation list $\pi_u$, user perceived experience $Z(\pi_u)$ comprises the utility $V$ and the regret $R$ based on a reference point $\pi'$.
In this paper, we set this reference point as the recommendation list, i.e., $\pi' = \pi_u^*$  with the maximized accuracy, as users always prefer to receive recommendations of higher quality. This regret $R$ is non-linear, meaning that the lower the recommendation quality, the higher the degree of regret.  In particular, when user accuracy $q_{\pi_u}$ is low, this regret can lead to a significant drop in the user’s perceived utility $Z({\pi_u})$, thus affecting their overall satisfaction. Therefore, to balance fairness objectives $G(\bm{e})$ with user experience $Z({\pi_u})$, the platform needs to model the user utility $V$ and the regret $R$ towards the recommendation list $\pi_u$.


$\bullet$ \textbf{Utility Function.}
The utility function of user shows the risk-aversion characteristics of human behavior. We also use a power function~\cite{loomes1982regrettheory} to represent the utility of user $u$ towards re-ranking list $\pi_u$:
\begin{equation}
    V\left(\pi_u\right) = q_{\pi_u}^\alpha= q_{\pi_u},
\end{equation}
where $0<\alpha \leq 1$ is the risk aversion coefficient of users. In recommendation scenarios, the risk users face due to high-quality recommendations is very low. Therefore, their risk aversion is not significant, and they always hope to receive the highest possible recommendation quality. Hence, in this paper, we make $\alpha=1$, which means users are risk-neutral.

$\bullet$ \textbf{Regret-rejoice Function.}
The regret-rejoice function indicates the user will experience improved regret when receiving a poor-quality recommendation compared to what was expected. We define the regret-rejoice function in re-ranking as:
\begin{equation}
\begin{aligned}
    R\left(\Delta V\right) & = R\left[V\left(\pi_u\right)-V\left(\pi^*_u\right)\right]\\
    &=  1- \exp\left(-\delta \left(V\left(\pi_u\right)-V\left(\pi^*_u\right) \right) \right)  \\
    &=  1- \exp\left(-\delta \left(q_{\pi_u}- q_{\pi_u}^{*} \right) \right),  \\
\end{aligned}
\end{equation}
where $\delta > 0$ is the regret avoidance coefficient of user, and the greater $\delta$ related to, the larger the regret avoidance degree of the user. 

$\bullet$ \textbf{Perceived Satisfaction Function.}
According to the regret theory~\cite{loomes1982regret}, after the user $u$ receives the ranking list $\pi_u$, the true satisfaction (i.e., perceived utility) of $u$ is composed of two parts: the utility value and the regret-rejoice value:
\begin{equation}
\begin{aligned}
      Z\left(\pi_u\right) &= V\left(\pi_u\right) + R\left[V\left(\pi_u\right)-V\left(\pi^*_u\right)\right] \\
    & =  q_{\pi_u} +  1- \exp\left(-\delta \left(q_{\pi_u}- q_{\pi_u}^{*} \right) \right) ,
\end{aligned}
\label{eq:perceived_satis}
\end{equation}
where $V\left(\pi\right)$ represents the utility value of list $\theta$ and $R(\cdot)$ is the regret-rejoice value.


\subsubsection{\textbf{Step 2: Balancing Individual Accuracy and Provider Fairness via Fuzzy Programming}}
\label{sec:step2}
After getting the non-linear user satisfaction function $Z(\pi(u))$, we then need to combine such non-linear function with the provider fairness function to conduct two-sided re-ranking ensure individual fairness.
To this end, we formally write a fuzzy programming problem with dual objectives of individual user satisfaction and provider fairness. Then, we convert it into a solvable optimization problem.

$\bullet$ \textbf{Two-sided Fair Re-ranking as Regret-aware Fuzzy Programming.}
\label{sec:fuzzy_problem}
To ensure individual user accuracy while considering provider fairness, an intuitive approach is to impose a constraint for each user. However, this approach has two drawbacks:
(1) Too many constraints may lead to difficulties in solving the problem, potentially resulting in no feasible solution. This significantly increases the complexity of the problem and reduces efficiency.
(2) In practice, during the re-ranking stage, a minimum user accuracy threshold is typically not explicitly set. Instead, the goal is for each user’s satisfaction level to be “as high as possible.”

 To address these two challenges, we adopt the concept of fuzzy optimization to solve this problem. By fuzzifying the original fairness and user perceived satisfaction objectives, we ensure that fairness does not excessively compromise accuracy in the form of rigid constraints.
We formulate the two-sided re-ranking problem as a fuzzy goal programming:
\begin{align}
\label{eq:obj1}
&Z\left( q_{\pi_u}\right) \gtrsim  Z\left( q_{\pi_u}^*\right),  \quad \forall u \in [1,2,\cdots, |\mathcal{U}|]\\
\label{eq:obj2}
&G(\bm{e}) \lesssim G(\bm{e}^*)\\
\label{eq:epbar}
\text { s.t. } \quad 
& \bm{e}_p = \frac{1}{\sum_k p(k) K T }\sum_{u=1}^ {|\mathcal{U}|}\sum_{i \in \mathcal{I}_p}\sum_{k=1}^K p(k)\bm{X}_{u,i,k}, \forall p \in \mathcal{P}\\
  & \sum_{i\in\mathcal{I}} \bm{X}_{u,i,k} = 1, \quad \forall t\in [1,2,\cdots,  |\mathcal{U}|], k \in [1,\cdots, K], 
  \\ & \sum_{k=1}^K \bm{X}_{u,i,k} \leq1, \quad \forall i \in \mathcal{I},u\in [1,2,\cdots,  |\mathcal{U}|], \\
  & \bm{X}_{u,i,k} \in \{0,1\}, \quad \forall u,i,k,
\end{align}
where $\bm{X}_u$ is the three dimension decision variable for user $u$. $\bm{X}_{u,i,k} =1$ means item $i$ being ranked at the $k$-th position for user $u_t$. 
$\bm{e}$ is the exposure vector considering the position of item in the ranking list\footnote{We denote $\Delta^\mathcal{X}$ as the $|\mathcal{X}|$-simplex, that is, the set $\Delta^\mathcal{X}= \{\bm{v}\in \mathbb{R}_{\geq0}^{|\mathcal{X}|}| \sum_i v_i = 1 \}$},
where each element $\bm{e}_p$ means the probability that provider $p$ being exposed at each rank position during the re-ranking process of $|\mathcal{U}|$. The objective of this fuzzy programming includes two parts:
\textbf{(1) Perceived Satisfaction fuzzy goal~\eqref{eq:obj1}} indicates that the recommendation accuracy for each user $u_t$ is approximately greater than or equal to the desired level $Z(q_{\pi_u}^*)$;
\textbf{(2) Fairness fuzzy goal~\eqref{eq:obj2}} requires that provider unfairness $G(\bm{e})$ is nearly less than the desired unfairness level $G(\bm{e}^*)$. 

$\bullet$ \textbf{Equivalent Form of Two-sided Fair Fuzzy Programming.} After formulating the re-ranking process as a fuzzy goal programming problem, our objective is to transform it into a computable optimization problem. 

\begin{theorem}
\label{theo:equa_form}
When altering the recommendation accuracy measurement $q_{\pi_u}$ as the  Discounted Cumulative Gain(DCG)~\cite{jarvelin2002ndcg, saito2022fairrankingasdivision}, 
and the provider fairness measurement as variance of $\bm{e}/\gamma$, i.e.,

 \begin{equation}
\left\{
\begin{aligned}
&q_{\pi_u} = \sum_{u=1}^ {|\mathcal{U}|}\sum_{i \in \mathcal{I}_p}\sum_{k=1}^K p(k)\bm{X}_{u,i,k} s_{u,i}, \forall u \in \mathcal{U}, \\
 &  G(\bm{e})= 
\text{Var}(\frac{\bm{e}}{\gamma}) = \frac{1}{n-1} \sum_{i=1}^{n} (\frac{\bm{e}_p}{\gamma_p} - \frac{1}{n} \sum_{i=1}^{n} \frac{\bm{e}_p}{\gamma_p})^2,  \\
\end{aligned}
    \right.
 \end{equation}
Based on the compromised approach~\cite{chung2018fuzzy},  Equation~\eqref{eq:oriprob} is the equivalent form of the fuzzy goal programming problem of Equation (15-21):

\begin{equation}
\begin{aligned}
\label{eq:oriprob}
    \max_{\bm{X}_t} \quad & \frac{1-\lambda}{|\mathcal{U}| }\sum_{u=1}^{|\mathcal{U}| }Z'(\pi_u) + 
    \lambda \cdot G'(\bm{e}) \\ 
    \text{s.t.} \quad
 &\text{Constraint (9-12)},
    \end{aligned}
\end{equation}

where 
\begin{equation}
\left\{
    \begin{aligned}
    &    Z'(\pi_u)  \frac{q_{\pi_u}- \exp(-\delta \left(q_{\pi_u}-q_{\pi_u^*})\right)+\exp(q_{\pi_u^*})}{q_{\pi_u}^* - 1+\exp(q_{\pi_u^*})}, \\
& G'(\bm{e}) =  1 - \frac{1}{1 + exp\left(-k\left(Var(\bm{e}) - \frac{ g_0}{2}\right)\right)}
    \end{aligned}
      \right.
\end{equation}
   where $\lambda$ is the user satisfaction-provider fairness trade-off hyperparameter. Higher $\lambda$ represents the platform’s higher emphasis on provider fairness. $g_0$ is the degree of unacceptable unfairness, which can be adjusted based on the platform requirement for provider fairness.
\end{theorem}


\subsubsection{\textbf{Regret-aware Online Learning Algorithm}}
In real online recommendation
scenarios, users usually arrive one after the other sequentially (see
Figure 4) and we should give user $u_t$ recommendation list $\pi_u$
immediately. Hence, we develop an online version of the offline
algorithm.  The workflow of our proposed two-sided online fair re-ranking algorithm is illustrated in Figure~\ref{fig:big}.

\begin{algorithm}[t]
        \caption{Online learning algorithm of BankFair+}
    	\label{alg:bankfair}
    	\begin{algorithmic}[1]
    	\REQUIRE Initial dual variable $\bm{\mu}$, step size $\eta_t$, item-provider matrix $\bm{A}$, fairness hyperparameter $\lambda$, unacceptable fairness degree $g$, desired exposure $\gamma$.

\ENSURE The re-ranking decision variable $\{\bm{X}_t, t=1,2,..., \infty\}$.
\FOR{$n=1$ to $N$} 
\STATE $// ~~\texttt{module 1}$.
\STATE $\mathcal{N}=\{n,n+1,\cdots,N \}$.
\STATE  $\hat{\bm{r}}(n)=g\left(\left[\bm{r}_{1}, \bm{r}_{2}, \cdots, \bm{r}_{n-1}\right]\right)$.
\STATE $\bm{D}_p(n)=\alpha K \hat{\bm{r}}(n)$.
\STATE  $\hat{\bm{m}}_{p}(n) = \left[ \hat{\bm{m}}_{p}(n-1) -  \bm{M}_{p,n-1}  + \bm{\beta}_p \right]_{+}, \forall p\in\mathcal{P}$.
\STATE $\hat{\bm{M}}_{p}(n) = \mathrm{TAL}(\mathcal{N},\hat{\bm{m}}_p(n),\bm{D}_p(n)), \forall p\in\mathcal{P}$.
\STATE 
$    \bm{M}_{n}^{\top}  =  \hat{\bm{M}}(n)_1$.
\STATE $// ~~\texttt{module 2}$.
\STATE Initialize dual variable $\boldsymbol{\mu}=0$.
\STATE Update remaining unearned exposure $\bm{\beta}_n^\top=\bm{M}_{n}^{\top}$.
\FOR{$t=1$ to $T$}
\STATE \textcolor{red}{\texttt{// step 1: measuring user perceived satisfaction}}.
\STATE Receive preference score of user $u_t$: $\bm{s}_{u_t}$.   
\STATE Compute accuracy-maximized utility $V(\pi_{u_t}^*)=q_{\pi_u}^{*} = \max_{\bm{X}_t \in \mathcal{X}_t} \{\bm{s}_t^\top \bm{X}_t \bm{p}\}$.
\STATE Get user regret: $R(\Delta V)=1 -\exp\left[\delta(\bm{s}_t^\top \bm{X}_t \bm{p} -V(\pi_{u_t}^*) )\right]$.
\STATE Get user perceived satisfaction: $Z(\pi_{u_t}) = V(\pi_{u_t})+ R(\Delta V)$.
\STATE Normalize the user perceived satisfaction: $Z'(\pi_{u_t}) = Norm(Z(\pi_u))$.
\STATE \textcolor{blue}{\texttt{// step 2: online recommendation and updating
}}.
\STATE Make recommendation:
$\bm{X}_{t}=\arg \max _{\bm{x}_t \in \mathcal{X}_t}\left\{ \frac{1-\lambda}{|\mathcal{U}|}Z'(\pi_{u_t})  -\bm{\mu}_{t}^{\top} \bm{A}^{\top} \bm{X}_t \bm{p}\right\}$

\STATE Compute target exposure: $\bm{e}_{t}=\arg \max _{\bm{e}'\in  \Delta_{+}^{\mathcal{P}}}\left\{G_{\bm{\mu}}^*(-\bm{\mu}_t)\right\}$
\STATE Compute subgradient $\bm{g}_t \in \mathbb{R}^{|\mathcal{P}|}$: $\bm{g}_{t}=\bm{e}_{t}-\bm{A}^{\top}\bm{X}_{t}\bm{p}$.
\STATE 
$\bm{\mu}_{t+1}=\arg \min _{\bm{\mu} \in \mathcal{D}}\left\langle\bm{g}_{t}, \bm{\mu}\right\rangle+\frac{1}{ \eta}\left\|\bm{\mu}-\bm{\mu}_{t}\right\|_{w}^{2}.$
\ENDFOR
\ENDFOR
    	\end{algorithmic}
    \end{algorithm}




$\bullet$ \textbf{Dual Problem.}
To efficiently solve Equation~\eqref{eq:oriprob}, we utilize Lagrangian relaxation,
transforming Equation~\eqref{eq:oriprob} into a simpler regularized optimization
problem utilizing the following Theorem.

\begin{theorem}
\label{theo:dual}
Equation~\eqref{eq:dual} is the dual problem of Equation~\eqref{eq:oriprob}
\begin{equation}
\label{eq:dual}
 Y_{Dual} =  \min _{\bm{\mu} \in \mathcal{D}} \left[Z^{*}\left(\bm{A}\bm{\mu} \right)\right]+G^{*}(-\bm{\mu}),
    \end{equation}
where $\mathcal{D}=\left\{\bm{\mu} \in \mathbb{R}^{|\mathcal{P}|} : -\infty<G^*(-\bm{\mu}) < + \infty\right\}$ is the feasible region of dual variable $\bm{\mu}$ and 
    \[Z^*(\bm{A}\bm{\mu}) = \max _{\bm{X}_{t} \in \mathcal{X}}\sum_{t=1}^{\bm{r}_n}\left[\frac{1-\lambda}{|\mathcal{U}|}Z'(\pi_u)  -\bm{\mu}^{\top}  \bm{A}^{\top} \bm{X}_{t} \bm{p}\right],
    \] 
    \[
    G^*(-\bm{\mu})=\max_{\bm{e} \in \Delta^\mathcal{P}}\left[ \lambda G'(\bm{e}) + \bm{\mu}^{\top} \bm{e}\right],
    \]
where $\bm{p} \in \mathbb{R}_{+}^K$ denotes the examine probability vector and $\bm{p}_k=p(k)=\frac{1}{\log_2(k+1)}$  
\end{theorem}

$\bullet$ \textbf{Online Learning Algorithm.}
Our online learning algorithm process is shown in Algorithm~\ref{alg:bankfair}, which is composed of two steps.
At the first step, when a user $u_t$ visits the RS, given the preference score vector $r_t$.
The algorithm first calculates the quality of the recommended list that the current user $u_t$ expects, which maximizes the preference scores of the items in the recommended list. Then, based on this recommendation quality $q^*_{\pi_{u_t}}$, we can calculate the utility $V^*(\pi_{u_t})$. After obtaining the utility $V^*(\pi_{u_t})$, we combine it with our decision variable $\bm{X}$ to calculate the user’s regret $R$ through the regret-rejoice function $R(\cdot)$. Finally, by combining utility and regret, we can obtain the user’s perceived satisfaction $Z(\pi_{u_t})$. To facilitate subsequent optimization, we need to normalize the user’s perceived satisfaction to $Z'(\pi_{u_t})$.

At the second step,
 the algorithm computes the optimal allocation $\bm{X}^*$ which maximizes an opportunity exposure-adjusted reward, based on the current dual solutions $\bm{\mu}_t$.
Finally, the target exposure $\bm{e}_t$ is computed by computing the value which maximizes the regularizer $\lambda G'(\bm{e})$ adjusted by an additive term accounting for the current dual solution $\bm{\mu}_t$.
First, the algorithm computes an unbiased stochastic estimator of a subgradient of $Y_{Dual}$ at $\bm{\lambda_t}$. The algorithm employs this estimator to update the vector of dual variables by performing an online dual mirror descent with step size $\eta$. 
Intuitively, at each time step $t$, the algorithm compares (i) the actual exposure from the recommendation to the expected exposure per iteration, and (ii) the target distribution over providers at $t$ to the realized provider. If the actual expenditure is higher (resp., lower) than this expected rate, then the algorithm surmises that future opportunities will offer higher (resp., lower) bang for the buck than current opportunities, and therefore the algorithm lowers (resp., raises) the bid shading multiplier. At the same time, if the realized providers cause an undesired skew, and move the realized distribution of impressions away from the desired  $\bm{e}_t$, then the algorithm will adjust its dual solutions to penalize allocations from that provider, and increase the likelihood of acceptance for items coming from under-exposed providers.


\section{Experiment}
\label{sec:experiment}
In this section, we empirically verify the effectiveness of BankFair+
by addressing the following research questions:
\\
$\bullet$ \textbf{RQ1}: How effective is BankFair+ in ensuring the user average accuracy and provider fairness?
\\
$\bullet$ \textbf{RQ2}: How effective is BankFair+ in ensuring the user individual fairness and provider fairness?
\\
$\bullet$ \textbf{RQ3}: What is the distribution of user accuracy after re-ranking with BankFair+, and is it fairer compared to the baseline?
\\
$\bullet$ \textbf{RQ4}: How does BankFair+ perform on different base ranking models?
\\
$\bullet$ \textbf{RQ5}: How do the hyperparameters affect the performance of BankFair+?

\subsection{Experimental Settings}

\subsubsection{Datasets}
Follow the previous work~\cite{ye2024bankfair}, we use two industrial recommendation datasets, which contain interactions between users, items, and providers, i.e., KuaiRand, and Huawei-Video.

$\bullet$ \textbf{KuaiRand-1K}\footnote{\url{https://kuairand.com/}}~\cite{gao2022kuairand}: a dataset collected from the video-sharing mobile app, \textit{KuaiShou}. The data are all derived from interactions from 8th April 2022 to 8th May 2022 from \textit{KuaiShou}.
We use the standard part, which contains randomly sampled user traffic of 302870 interactions from 933 users on 6825 videos with 174 providers.

$\bullet$ \textbf{Huawei-Video}: a dataset collected from the short video platform in \textit{Huawei Browser} from 2 Jan. 2024 to 8 Jan. 2024, spanning a week. It contains 19355 users, 5364 items, and 200 providers, totaling 118765 interactions. 

For KuaiRand-1K, we use the interactions before 23rd April for training and 23rd April to 8th May for testing. For Huawei-Video, we use interactions before 5th Jan. for training and 6th Jan. to 8th Jan. for testing.

\subsubsection{Baselines}
We compare with the following provider fairness-aware baselines: 

$\bullet$ \textbf{BankFair}: online re-ranking algorithm that first utilizes bankruptcy problem to allocate minimum exposure to providers.

$\bullet$ \textbf{Welf}~\cite{do2021welf}: use the Frank-Wolfe algorithm to maximize the Welfare functions of worst-off items in an offline manner.

$\bullet$ \textbf{P-MMF}~\cite{xu2023p}: an integer-programming (IP)-based re-ranking method ensures the exposure of worst-off provider; 

We also compare some works that consider two-sided fairness, i.e., ensuring user individual fairness while maintaining provider fairness:

$\bullet$ \textbf{CPFair}~\cite{naghiaei2022cpfair}: an efficient greedy algorithm capable of achieving an optimal trade-off within between user and provider; 

$\bullet$ \textbf{FairRec}~\cite{patro2020fairrec}: a method guarantees the minimum exposure for providers and employs a greedy strategy to uphold user fairness; 

$\bullet$ \textbf{TFROM}~\cite{wu2021tfrom}: A two-sided re-ranking method which improves provider exposure and amortizes the accuracy loss among users to guarantee individual fairness; 

$\bullet$ \textbf{PCT}~\cite{wang2023pct}: A max-margin-relevance-based method to ensures a target item exposure distribution and reduce accuracy loss for users.

\subsubsection{Evaluation Metrics} 

For provider fairness evaluation, we utilize two widely-used fairness metric, i.e, enough satisfaction group (ESP)~\cite{ye2024bankfair} and Gini Index~\cite{do2022optimizegini}. Meanwhile, different from previous work, we considered the position bias in the exposure (see Equation~\ref{eq:exposure_with_decay}), which is a crucial factor that influences user examine probability and is overlooked by previous work~\cite{ye2024bankfair}.

For the ESP metric, we aims to evaluate the capability that different methods can guarantees a minimum exposure for each provider. The ESP metric measures the provider proportion that receives an exposure value no less than the minimum exposure guarantee $\bm{m}$, i.e.,
the metric of ESP~\cite{wang2023uncertainty,patro2020fairrec, xu2024fairsync}:
\begin{equation}
\begin{aligned}
\mathrm{ESP@K} & =\frac{1}{|\mathcal{P}|} \sum_{p \in \mathcal{P}} \mathbb{I}\left(\bm{e}_p\geq \bm{m}\right)  \\ 
&= \frac{1}{|\mathcal{P}|} \sum_{p \in \mathcal{P}} \mathbb{I}\left(\left[\sum_{u\in\mathcal{U}}\sum_{i \in \pi(u,K)}p(\sigma(i)) \cdot \mathbb{I}(i\in \mathcal{I}_p)\right]\geq \bm{m}_{p}\right).
\end{aligned}
\end{equation}

 where $\mathbb{I}(\cdot)$ is the indicator function and $p(\cdot)$ is the position decay function. In this experiment, we define with $p(k):=\frac{1}{\log_2(k+1)}$.  The value of ESP ranges between 0 and 1. When all the providers get exposures no less than the minimum exposure guarantee $\bm{m}_p$, then ESP becomes 1.

Besides the ESP metric which focuses on guaranteeing the minimum exposure for providers, we also use the Gini Index~\cite{do2022optimizegini} to assess whether each provider can receive exposure $\bm{e}_p$ comparable to their merit $\bm{\gamma}_p$. The detailed mathematical expression is defined as follows:

\begin{equation}
\text { Gini}@K=\frac{\sum_{p_1, p_2 \in \mathcal{V}}\left|\frac{\bm{e}_{p_1}}{\bm{\gamma}_{p_1}}-\frac{\bm{e}_{p_2}}{\bm{\gamma}_{p_2}}\right|}{2|\mathcal{P}| \sum_p \frac{\bm{e}_{p}}{\bm{\gamma}_{p}}}
\end{equation}

The value of Gini ranges between 0 and 1. When all the providers get exposures proportional to their merit $\bm{r}_p$, then $\text{Gini}@K$ becomes 0. The higher Gini is, the more unfairness providers experience.


For user accuracy, we use NDCG~\cite{jarvelin2002ndcg} as our user accuracy metric. The specific definition of NDCG is defined as follows.
\begin{equation}
  \mathrm{NDCG}_{u_t}@\mathrm{K} = \frac{\sum_{i \in \pi (u, K)} s_{u, i} / \log \left(\operatorname{rank}_{i}+1\right)}{\sum_{i \in \pi^*(u, K)\left(u\right)} s_{u, i} / \log \left(\operatorname{rank}_{i}^{\text{ori}}+1\right)}  
\end{equation}
$\text{rank}_i$ and $\text{rank}_i^{\text{ori}}$ are the ranking positions of the item $i$ in $\pi (u, K)$ and $\pi^* (u, K)$.  

For user individual fairness, to measure the accuracy gap between users, we use Min-Max Ratio (MMR) as one metric to measure user 
 individual unfairness. The greater the MMR value, the greater the user individual unfairness.
\begin{equation}
\text {MMR}@K=\frac{\min \{\text{NDCG}_{u_1}@K, \forall u_1 \in \mathcal{U}\}}{\max \{\text{NDCG}_{u_2}@K, \forall u_2 \in \mathcal{U}\}}
\end{equation}

 The value of MMR ranges between 0 and 1. When all the users get the same accuracy, then ESP becomes 1. Otherwise, the greater the unfairness in the NDCG across different individual users, the smaller the MMR will be.

\subsubsection{Implemental Details}
Following previous works~\cite{xu2023p}, the merit of each provider is set to the sum of the relevance scores of the items they own: $\bm{\gamma}_p = \frac{|\mathcal{I}_p|}{|\mathcal{I}|}$. 
In this paper, instead of setting the same minimum exposure guarantee for every provider~\cite{ye2024bankfair, xu2024fairsync} (i.e., $\bm{m}_{p_1}=\bm{m}_{p_2}, \forall p_i, p_j\in \mathcal{P}$), we choose different minimum exposure guarantees for each provider. Specifically, we provide each provider with a minimum exposure guarantee based on their merit, i.e., $\bm{m}_p = \beta \frac{\bm{\gamma_p}}{\sum_p \bm{\gamma}_p}\sum_{u\in\mathcal{U}}\sum_{i \in \pi(u,K)}p(\sigma(i))$. In the main experiment, we set $\beta = 0.9$, which means that each provider cannot receive less than 90\% of the exposure they are supposed to receive.

\subsection{Experimental Result}

\begin{figure}[ht]
    \centering
   \centering
        \begin{subfigure}{\linewidth}
	\centering
	\includegraphics[width = \linewidth]{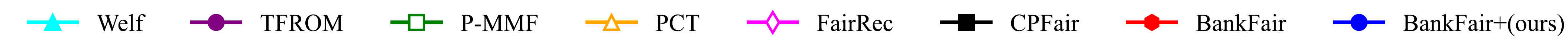}
	\end{subfigure}
    \begin{subfigure}{\linewidth}
	\centering
	\includegraphics[width = \linewidth]{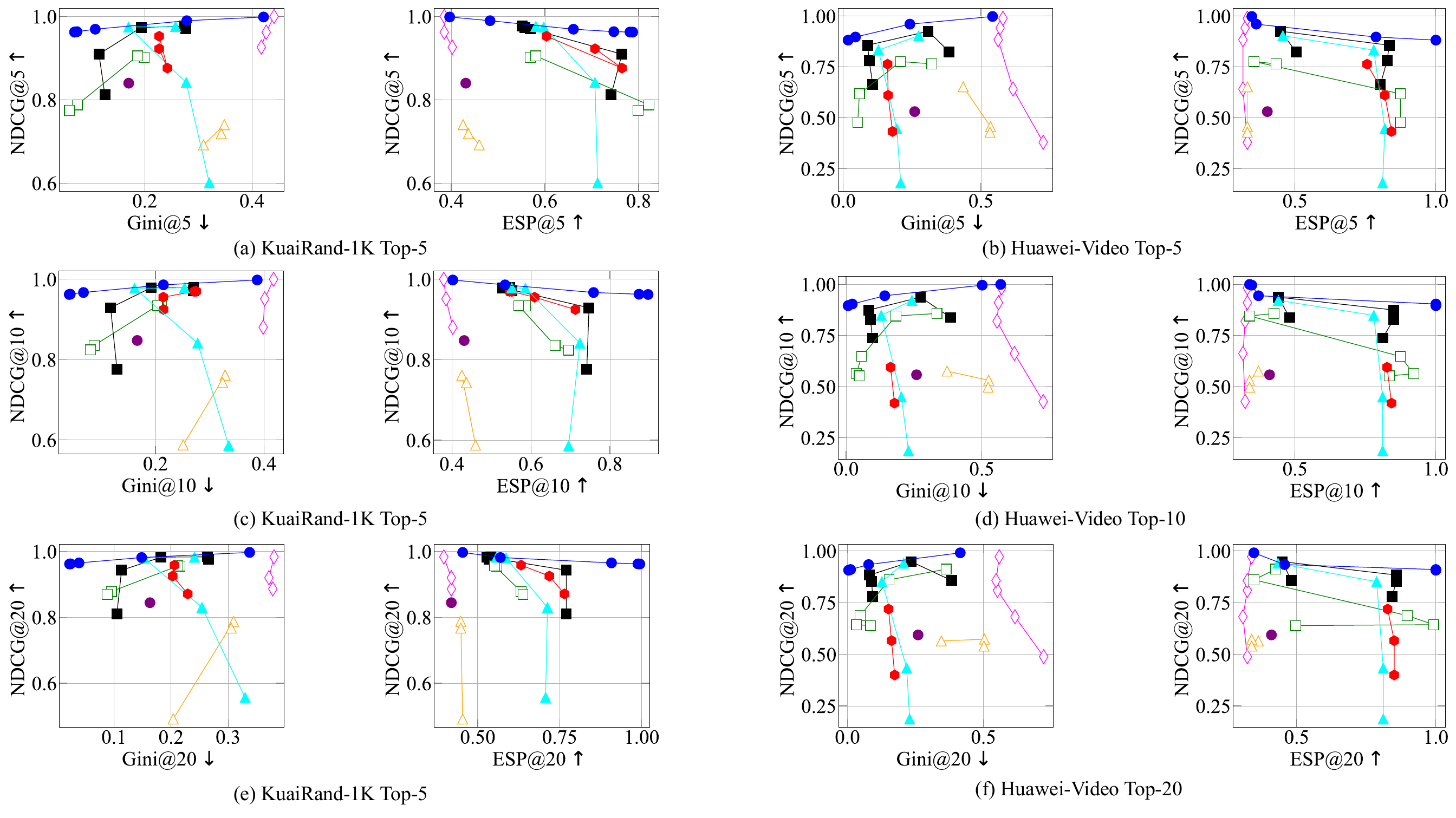}
	\end{subfigure}
    \caption{The Pareto frontierof user accuracy (NDCG@K) and provider fairness (Gini@K and ESP@K) on two different datasets with top-$10$. X-axis shows Gini@K metric and ESP@K metric, while Y-axis shows NDCG@K metric. $\uparrow$ means higher values are better and $\downarrow$ favors lower values.}
    \label{fig:pareto_avg_acc}
\end{figure}

In this section, we conduct experiments to test the performance of BankFair and baselines in terms of user accuracy, user individual unfairness, and provider fairness, on two different datasets and Huawei-Video datasets. By adjusting the hyper-parameter that controls the degree of fairness, we can adjust the degree of the trade-off between
accuracy and provider fairness.  For a fair
comparison between baselines, we visualize the change curves
between metrics and comprehensively compare different methods
based on the concept of Pareto efficiency~\cite{lotov2008pareto}. As shown in Figure~\ref{fig:pareto_avg_acc} and Figure~\ref{fig:pareto}, we
can see that our proposed BankFair+ achieves Pareto optimality over
baselines in terms of average accuracy, individual user fairness and two-sided fairness.

\subsubsection{\textbf{RQ1: Performance of guaranteeing average accuracy and provider fairness}}
We conduct experiments to verify the effectiveness of 
 BankFair+ in balancing the trade-off between the average user accuracy and provider fairness on both industrial recommendation datasets.
 As illustrated in Figure~\ref{fig:pareto_avg_acc}, we plot the Pareto frontiers of NDCG@K-Gini@K and NDCG@K-ESP@K on two datasets with different ranking sizes $K= 5, 10, 20$. The Pareto frontiers are drawn by tuning different parameters of the models and choosing the (NDCG@K, ESP@K)  points with the best performances.
The Pareto frontiers are drawn by tuning different parameters of the models and choosing the (NDCG@K, ESP@K) and (Vio@K, ESP@K) points with the best performances.

From the Pareto frontiers, we find that BankFair+ Pareto dominates the baselines (i.e., the BankFair+ points are at the upper left corner in the NDCG@K-Gini@K and at the upper right corner NDCG@K-ESP@K in Figure~\ref{fig:pareto_avg_acc}, which means BankFair+ achieves higher accuracy level (NDCG@K) under different levels of provider fairness (i.e., Gini@K and ESP@K).

At the same time, we also observe that, regardless of hyper-parameter adjustments, most baselines are unable to achieve very high provider fairness levels (with Gini@K below 0.1 and ESP@K above 0.8), especially for higher ranking sizes such as $K=10$ or $20$. However, in comparison, our proposed BankFair+ achieves a higher level of fairness, and at this higher level of fairness, we are still able to maintain a high average user accuracy. Specifically, as shown in the NDCG@10-Gini@10 curve plot in Figure~\ref{fig:pareto_avg_acc}, at the same level of fairness (Gini< 0.1), BankFair+’s NDCG is at least 0.2 higher than the best-performed baseline (i.e., P-MMF).
In the NDCG@10-ESP@10 curves in Figure~\ref{fig:pareto_avg_acc} and, at the same level of provider satisfaction (ESP=80\%), BankFair+’s NDCG is 0.18 higher than the best-performed baseline.


\subsubsection{\textbf{RQ2: Performance of guaranteeing user individual fairness and provider fairness}}
\label{sec:individual fairness and provider fairness}
\begin{figure}[ht]
    \centering
   \centering
        \begin{subfigure}{\linewidth}
	\centering
	\includegraphics[width = \linewidth]{fig/legend.pdf}
	\end{subfigure}
    \begin{subfigure}{\linewidth}
	\centering
	\includegraphics[width = \linewidth]{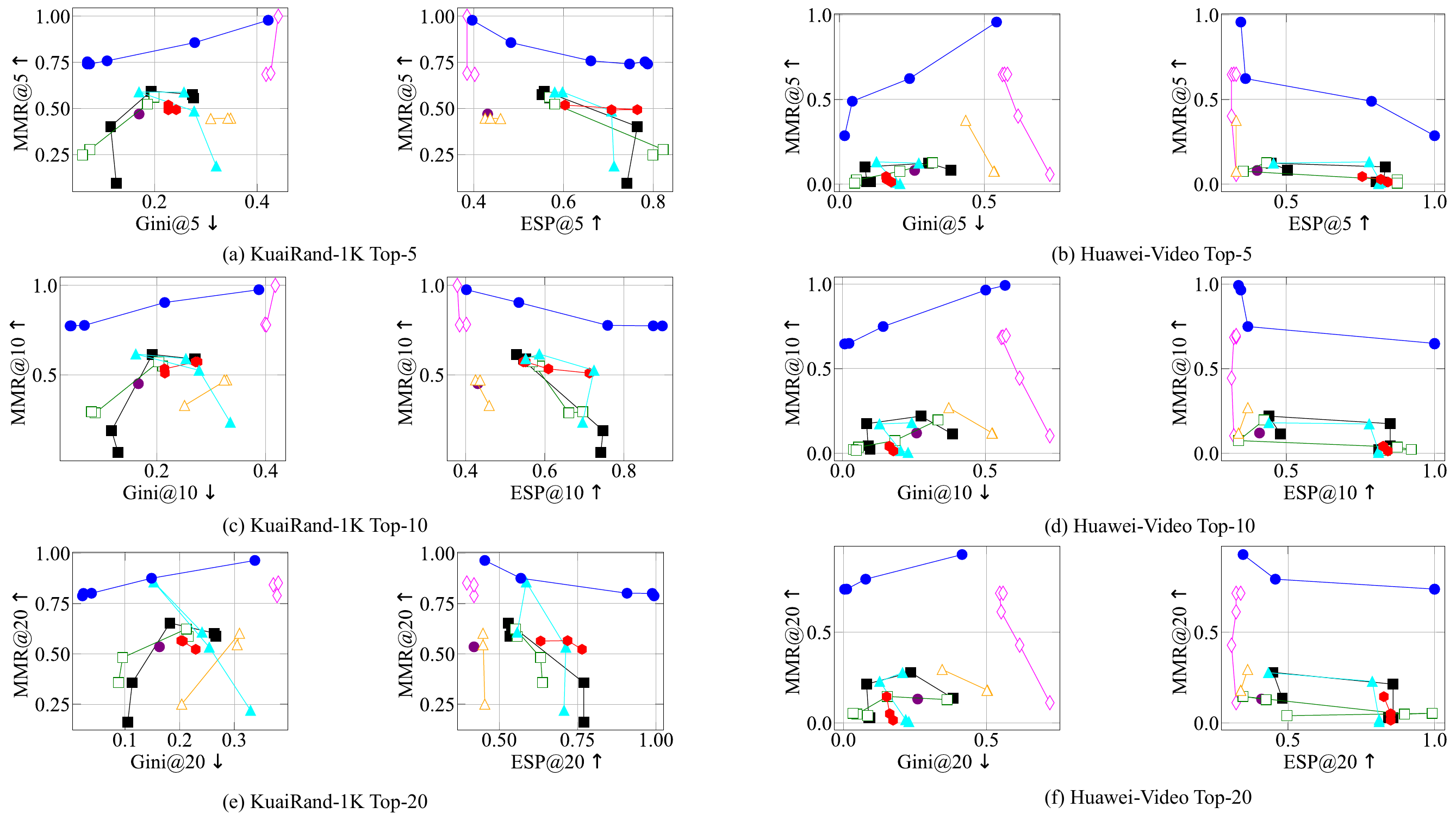}
	\end{subfigure}
    \caption{The Pareto frontier of individual accuracy (MMR@K) and provider fairness (Gini@K and ESP@K) on two different datasets with top-$10$. X-axis shows Gini@K metric and ESP@K metric, while Y-axis shows MMR@K  metric. $\uparrow$ means higher values are better and $\downarrow$ favors lower values.}
    \label{fig:pareto_ind_fair}
\end{figure}

We also conduct experiments to verify that our method can ensure both provider fairness and individual user fairness on different datasets. Figures~\ref{fig:pareto_ind_fair} illustrate the Pareto fronts of individual user fairness and provider fairness levels. Specifically, Figures~\ref{fig:pareto_ind_fair},  show the Pareto frontiers between MMR@K-Gini@K and MMR@K-ESP@K.


From Figure~\ref{fig:pareto_ind_fair}, we observe BankFair+, effectively ensures both individual user fairness and provider fairness. By adjusting the hyperparameters, our method can achieve Pareto-optimal individual user fairness at different provider fairness levels. Specifically, when Gini@K decreases from 0.4 to 0 and ESP increases from 0.4 to 0.8, MMR remains above 0.7. This means that users with the lowest recommended accuracy receive more than 70\% of the accuracy compared to users with the highest recommended accuracy. At the same provider fairness level, for instance when Gini=0.1 and ESP=0.7, the MMR metric of baselines does not exceed 50\%.



\subsection{Experimental Analysis}
We conducted experiments to analyze BankFair on KuaiRand-1K on ranking size $K=10$. In this section, we select the best-performed baselines in Section~\ref{sec:experiment} for comparison, i.e., BankFair, CPFair, and Welf.

\subsubsection{\textbf{RQ3: Case study of individual user accuracy distribution}}

\begin{figure}[!ht]
    \centering
    \includegraphics[width=1\linewidth]{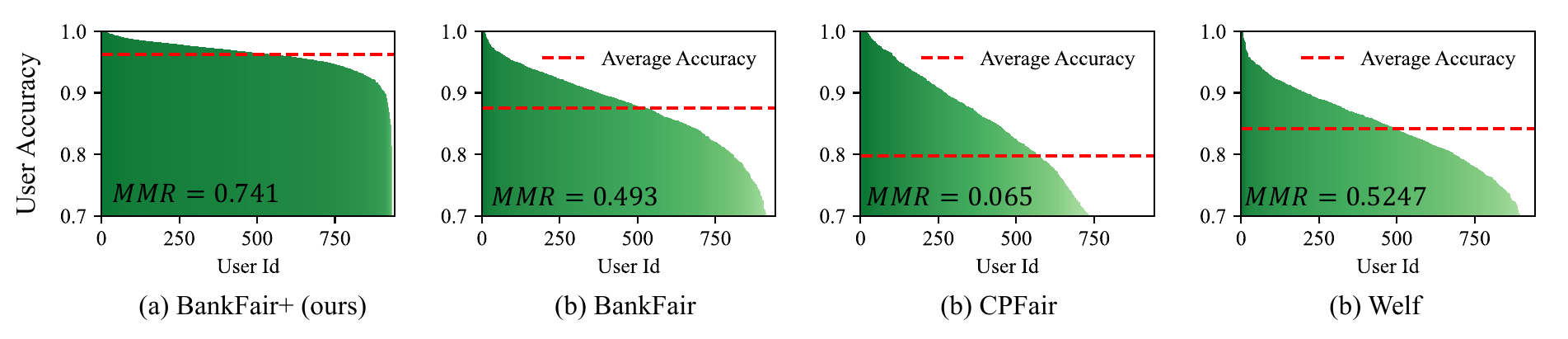}
    \caption{The user distribution on individual accuracy of our method BankFair+ and the best-performed baselines. The hyperparameters of the method were tuned to achieve the best average user accuracy with 75\% ESP@10. The red line indicates the average accuracy level of users.}
    \label{fig:user_distribution}
\end{figure}

To demonstrate the ability of BankFair+ in ensuring individual fairness for users, we plot the distribution of individual user accuracy for our method and the best-performing baselines. For a fair comparison, we tune the hyperparameters of the selected methods to achieve the best average user accuracy with at least 75\% ESP@10. As illustrated in Figure~\ref{fig:user_distribution}, we plot the distribution of the recommendation accuracy for the 933 users in KuaiRand-1K, sorted by user ID from highest to lowest accuracy. The red line represents the average accuracy of all users. The text “MMR” represents the value of the Minmax ratio.
We can see that our method is more individually fair than the baselines in terms of user recommendation accuracy. Specifically, the user distribution of the baselines is more long-tailed with higher variance, and the lowest user accuracy is much lower than the highest (the MMR of BankFair is 49.3\%, and CPFair is only 6.5\%). Meanwhile, when ensuring individual fairness, our method also achieves a higher average accuracy, indicating that our approach effectively balances individual fairness, average accuracy and provider fairness.

\subsubsection{\textbf{RQ3: Influence of different base model}}

\begin{figure}  
    \centering
            \begin{subfigure}{\linewidth}
	\centering
	\includegraphics[width = \linewidth]{fig/legend.pdf}
	\end{subfigure}
    \begin{subfigure}{\linewidth}
	\centering
\includegraphics[width=1\linewidth]{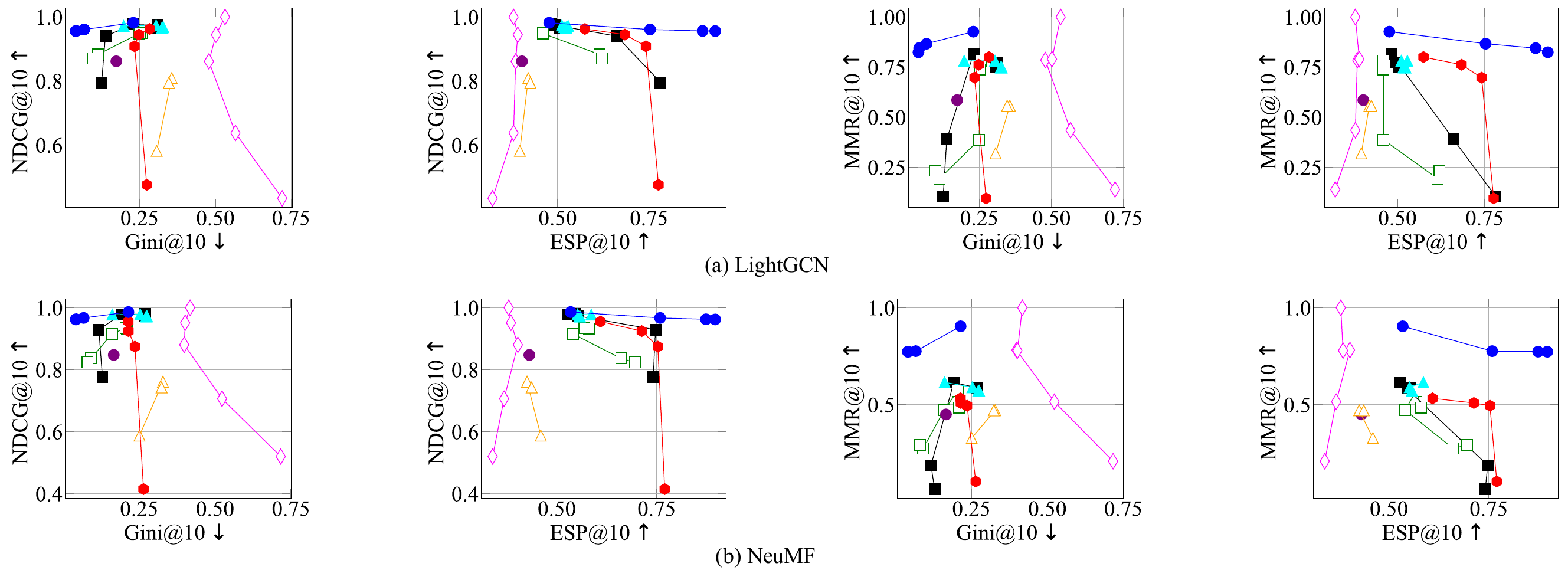} 
	\end{subfigure}
    \caption{Performance of BankFair+ and the baselines using different base ranking models. The Pareto frontier of the methods on different base ranking models (LightGCN, NeuMF) with ranking sizes $K=10$. X-axis shows Gini@K metric and ESP@K metric, while Y-axis shows NDCG@K, MMR@K metric. $\uparrow$ means higher values are better and $\downarrow$ favors lower values.}
    \label{fig:base_model}
\end{figure}


We conduct experiments on different base ranking models to demonstrate the effectiveness of our method. Figure~\ref{fig:base_model} shows the performance of our method and the baselines when using LightGCN~\cite{he2020lightgcn} and NeuMF~\cite{he2017neuralmf} as ranking models. We observe that our method effectively balances both average user accuracy and individual fairness for users under varying levels of vendor fairness. Specifically, in Figures~\ref{fig:base_model} (a) and (b), our method is Pareto superior to the other baselines. This indicates that our method can achieve two-sided fairness, ensuring fairness for both providers and users, while not sacrificing too much recommendation accuracy.

\subsubsection{\textbf{RQ4: Influence of different required minimum exposure threshold $\beta$}}

\begin{figure}[!ht]
    \centering
    \begin{subfigure}{\linewidth}
	\centering
	\includegraphics[width = 0.5\linewidth]{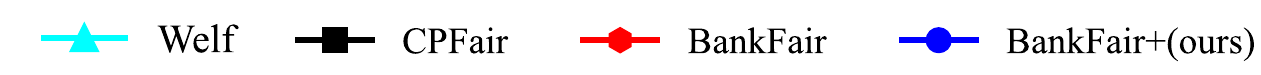}
	\end{subfigure}
    \begin{subfigure}{\linewidth}
	\centering
\includegraphics[width=1\linewidth]{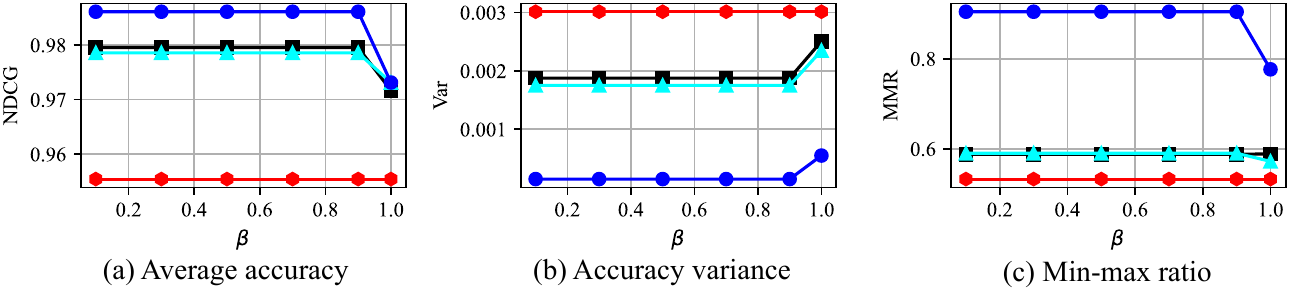}
	\end{subfigure}
    \caption{Performance of average accuracy and inidividual fairness w.r.t. minimum exposure threshold that controls the minimum exposure  
 $\bm{m_p}$ for each provider $p$.}
    \label{fig:threshold}
\end{figure}

In real applications, the platforms often change the required minimum exposure $\bm{m}$ to fit different applications (e.g., development stage, incentive policy)~\cite{bardhan2022more}. 
In this experiment, we investigate the performance of BankFair+ as the required minimum exposure $\bm{m}_p$ varies. In this experiment, we use $\beta$ to control the minimum exposure for each provider. Specifically, the relationship between $\bm{m}_p$ and $\beta$ is $\bm{m}_p = \beta \cdot \text{metrit}_p= \frac{\bm{\gamma_p}}{\sum_p \bm{\gamma}_p}\sum_{u\in\mathcal{U}}\sum_{i \in \pi(u,K)}p(\sigma(i))$, where $\text{merit}$ is the exposure that a provider $p$ should receive based on the size and quality of their owned items.
To make fair comparisons for all selected models, we select the result with the best NDCG@K performance with at least $\%$ ESP@K under different $\bm{m}_p$.

From NDCG@10, MMR@10, and Var@10 curves in Figure~\ref{fig:threshold}, we can see that the proposed BankFair+ achieves better average accuracy (higher NDCG@10) in comparison to the baselines under different required minimum exposure $\bm{m}_p$ levels. 
Moreover, BankFair can maintain relatively stable and better individual fairness (higher MMR@10 and lower Var@10) under different required minimum exposure $\bm{m}_p$ levels.

\subsubsection{\textbf{RQ5: Influence of regret avoidance coefficient $\delta$.}}

\begin{figure}
    \centering
    \includegraphics[width=0.9\linewidth]{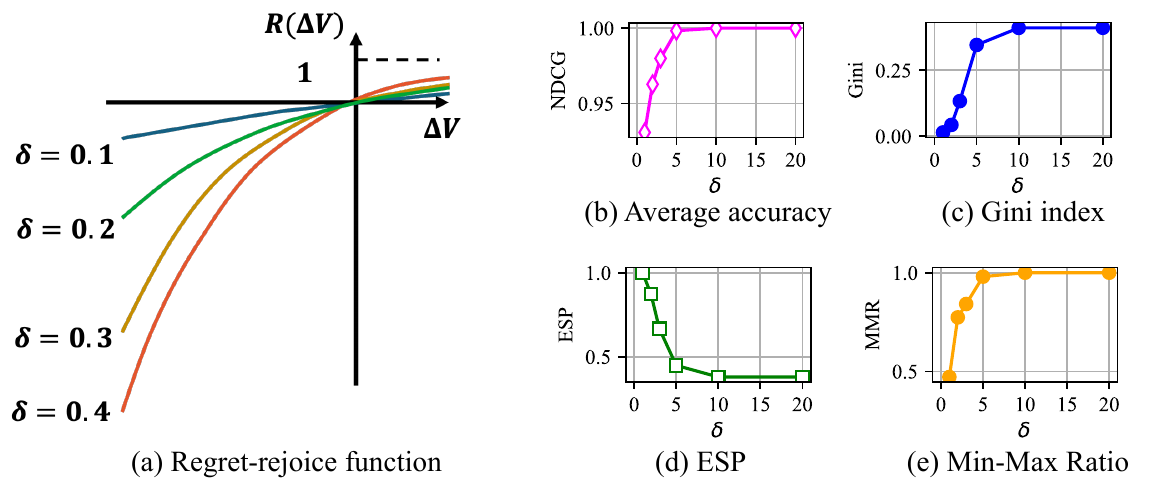}
    \caption{ (a) Regret-rejoice function w.r.t different regret avoidance coefficient $\delta$. Performance of (b) user accuracy, (c-d) provider fairness, and (e) user individual fairness with varying regret avoidance coefficient $\delta$. }
    \label{fig:sigma}
\end{figure}

In this part, we conduct experiments to explore how the performance of BankFair+ changes with the adjustment of the hyper-parameters $\delta$ related to the regret theory.  
Figure~\ref{fig:sigma} (a) shows the image of the regret-rejoice function with different regret avoidance coefficient $\delta$ values. 
In this experiment, the accuracy-fairness trade-off parameter is set to be $\lambda = 1000$.
Figure~\ref{fig:sigma} illustrates the performance of BankFair+ in terms of user accuracy (b), provider fairness (c-d), and individual user fairness (e) under different regret avoidance coefficient ($\delta$) values. 


From~\ref{fig:sigma} (b), we can observe that as $\delta$ increases, the average accuracy continuously rises and eventually approaches its maximum value. This is because, according to the regret theory (see Section~\ref{sec:regret_theory}) Equation~\eqref{eq:perceived_satis}, as $\delta$ increases, users become increasingly averse to regret and the proportion of regret $R$ in the user’s perceived satisfaction $Z(\pi_u)$ grows larger. Moreover, since regret $R$ decreases as satisfaction declines and given its first derivative $R'(\cdot)>0$ is positive and its second derivative $R''(\cdot)<0$ is negative, the rate of decline accelerates as the quality of the recommendation list $q_{\pi_u}$ deteriorates.

Consequently, during optimization, a lower recommendation quality $q_{\pi_u}$ for users inflicts greater damage on the optimization objective. As a result, the optimization process gradually ensures that all users receive similar higher recommendation quality $q_{\pi_{u_1}}\approx q_{\pi_{u_2}}, \forall u_1=u_2$. This explains why, as shown in Figure~\ref{fig:sigma} (e), individual user accuracy becomes increasingly fair (MRR metric increases with the increase of $\delta$). However, this improvement comes at the cost of some compromise in provider fairness (ESP metric decreases and Gini metric increases with the increase of $\delta$ in Figure~\ref{fig:sigma} (c-d)).

Nevertheless, we emphasize that although enhancing individual user fairness leads to a decline in provider fairness, our method still achieves higher provider fairness at the same level of individual user fairness, as evidenced by Section~\ref{sec:individual fairness and provider fairness}.

\section{Conclusion  and Future Works}
In this paper, we emphasize the importance of ensuring individual fairness while guaranteeing provider fairness and user accuracy in multi-stakeholder platforms. 
To mitigate the inidividual unfairness of the existing approaches, we propose an online two-sided re-ranking model, named BankFair+. By introducing regret theory from economics into the re-ranking phase of recommender systems, we introduce a non-linear user satisfaction function to balance user accuracy and individual fairness.
Furthermore, we propose an online regret-aware fuzzy programming algorithm to integrate provider fairness objective with individual user accuracy, thereby guaranteeing individual fairness and provider fairness simultaneously. 
Extensive experiments demonstrate BankFair+'s effectiveness in balancing individual fairness, provider fairness, and user accuracy.

As for the future work, this work points to new research possibilities. Currently, we aim to ensure individual fairness during the re-ranking stage. In the future, we plan to extend our approach to other stages of the recommendation system, such as the ranking stage. We will explore the relationships between provider fairness, user accuracy, and individual fairness in these stages.
\newpage
\appendix
    \section*{Appendix}

\section{Proof of Theorem~\ref{theo:equa_form}}

\begin{proof}
Before we write the equivalent form of Equation~\eqref{eq:oriprob}, we first need to introduce fuzzy logic to fuzzify the fairness constraints and accuracy objectives. Previous research~\cite{zimmermann1978fuzzy} has shown that this approach can minimize the negative impact on the objectives.

\begin{figure}[!ht]
    \centering
\includegraphics[width=1\linewidth]{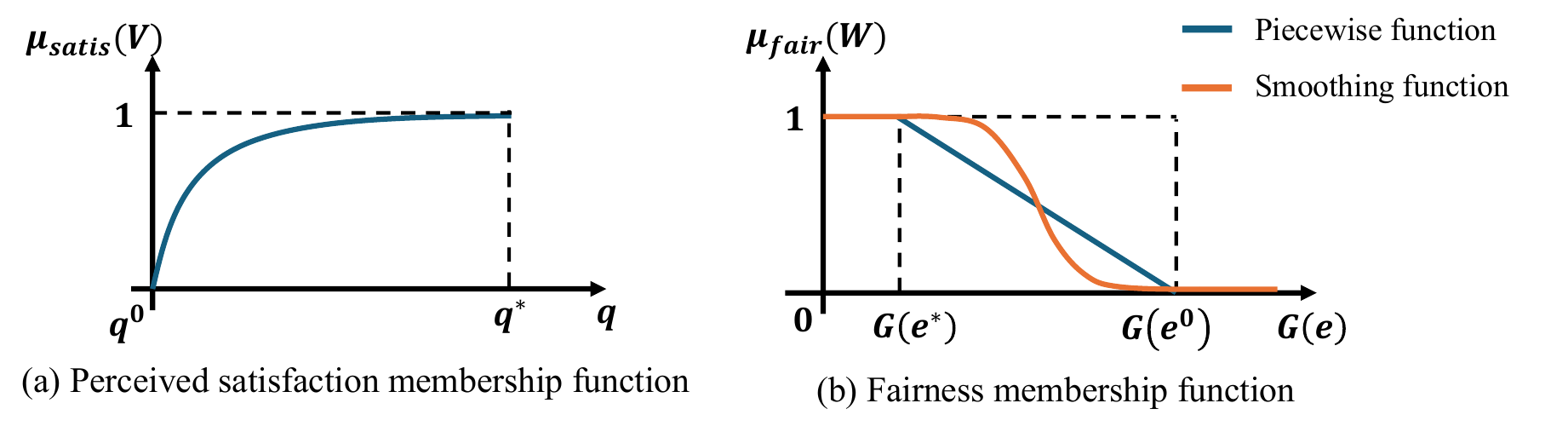}
    \caption{(a) Perceived satisfaction membership function; (b) fairness membership function.}
    \label{fig:member}
\end{figure}

\textbf{Membership Function of User Perceived Satisfaction.}
We first use fuzzy logic to represent user satisfaction, transforming it into a function of recommendation quality $q_{\pi_u}$ that maps to a range between 0 and 1. The member ship function of user-perceived satisfaction is formulated as follows:
\begin{equation}
\mu^{\text{satis}}\left(q_{\pi_u}\right)= 
\begin{cases}
   Norm\left(Z(\pi_u) \right) =  \frac{Z(\pi_u) - Z(\pi_u^0)}{Z(\pi_u^*) - Z(\pi_u^0)}, &  q_{\pi_u^0}< q_{\pi_u} \leq q_{\pi_u^*}
    \\
    0, & \text{otherwise}
\end{cases}
\label{eq:muz}
\end{equation}
where $Z(\cdot)$ is the user-perceived satisfaction function defined in Equation~\eqref{eq:perceived_satis}. $\pi_u^*$ and $\pi_u^0$ are the quality-maximized and quality-minimized recommendation lists, respectively.
The specific shape of the membership function $\mu^{\text{satis}}\left(q_{\pi_u}\right)$ is shown in Figure~\ref{fig:member} (a).
As the quality of the recommended list $q_{\pi_u}$ gradually increases to the maximum value $q_{\pi_u^*}$, the membership function $\mu^{\text{satis}}\left(q_{\pi_u}\right)$ also gradually rises to 1, indicating that the user’s perceived satisfaction increases and eventually reaches full satisfaction. Conversely, when the quality of the recommended list $q_{\pi_u}$ is low (i.e., approaching the minimum level $q_{\pi_u^0}$), the value of the membership function approaches 0, meaning the user’s perceived satisfaction $\mu^{\text{satis}}\left(q_{\pi_u}\right)$is low, or even close to complete dissatisfaction. In this process, the membership function changes in a non-linear manner (i.e., $\mu^{\text{satis}'}(\cdot) > 0$, $\mu^{\text{satis}''}(\cdot) < 0$) to better reflect the relationship between recommendation quality and user satisfaction. Figure~\ref{fig:member} (a) illustrate the value of the membership function $\mu^{\text{satis}}$ of user perceived satisfaction as the quality of the recommendation list $q_{\pi_u}$ changes.


\textbf{Membership Function of Fairness.}
We also use the membership function to characterize the fairness level, which uses a piecewise linear function:
\begin{equation}
\mu^{\text{fair}}\left(\bm{e}\right)= \begin{cases}0, & \text { if } G(\bm{e})\geq G(\bm{e}^*) ,\\-\frac{G(\bm{e})-G(\bm{e}^0)}{G(\bm{e}^*)- G(\bm{e}^0)}, & \text { if }  G(\bm{e}^0)\leq G(\bm{e}) \leq G(\bm{e}^*) ,\\
1, &\text{if } G(\bm{e}) \leq G(\bm{e}^0)\end{cases}
\end{equation}
where $\bm{e}$ is the exposure vector of providers. $\bm{e}^0$ is the undesired exposure value of providers, which means it is unacceptable if the unfairness level $G$ exceeds $G(\bm{e}^0)$. Similarly, $\bm{e}^*$ is the desired level, which means it is absolutely fair, i.e., $\bm{e}^*_{p_1}=\bm{e}^*_{p_2}, \forall p_1,p_2\in \mathcal{P}$.
Due to the non-differentiability of this piecewise function, which is not suitable for optimization, we adopt an S-shaped (sigmoid) function to replace it:
\begin{equation}
   \mu^{\text{fair}}\left(\bm{e}\right) = Sigmoid \left(G(\bm{e})\right) =  1 - \frac{1}{1 + exp\left(-k\left(G(\bm{e}) - \frac{G(\bm{e}^*) + G(\bm{e}^0)}{2}\right)\right)}.
\end{equation}
By choosing an appropriate value for $k$, we make this function closer to the original piecewise function.

After defining the membership function of both indiviudal user accuracy and provider fairness, we now formulate the fuzzy programming problem~\eqref{eq:oriprob} into a optimization problem.
According to the compromised approach~\cite{zimmermann1978fuzzy}, the above fuzzy programming can be expressed  as the following convex problem:
\begin{equation}
\begin{aligned}
\label{eq:oriprob}
    \max_{\bm{X}_t} \quad f & = \frac{1-\lambda}{|\mathcal{U}|}\sum_{u=1}^{|\mathcal{U}| }\mu^{\text{satisfy}}\left(Z(q_{\pi_u})\right) + 
    \lambda \cdot \mu^{\text{fair}}\left(G(\bm{e})\right)  \\ 
   \quad  &=  \frac{1-\lambda}{|\mathcal{U}|}\sum_{u=1}^{|\mathcal{U}| } \frac{Z(\pi_u) - Z(\pi_u^0)}{Z(\pi_u^*) - Z(\pi_u^0)} + 
    \lambda \cdot \left(  1 - \frac{1}{1 + exp\left(-k\left(G(\bm{e}) - \frac{G(\bm{e}^*) + G(\bm{e}^0)}{2}\right)\right)} \right) \\ 
    \text{s.t.} \quad 
 &\text{Constraint (9-12)},
    \end{aligned}
\end{equation}
where $\lambda$ is the user satisfaction-provider fairness trade-off hyperparameter. 

By choosing DCG~\cite{jarvelin2002ndcg, saito2022fairrankingasdivision} as the $p_{\pi_u}$ and variance $Var(\bm{\bm{e}/\gamma})$ as $G(\bm{e})$, we can know that   $G(\bm{e}*)=0$. Since the item set $|\mathcal{I}|$ is large and most items $i \in\mathcal{I}$ have a low relevance $s_{u,i}$ with the user, we can approximate that $q_{\pi_u}^0=0$. From Equation~\eqref{eq:perceived_satis}, we know that $Z({\pi_{u}}^0)=1-\exp(q_{\pi_u^*})$, $Z({\pi_{u}}*)=q_{\pi_u}^*$. After replacing the $Z({\pi_{u}}*), G(\bm{e}*)$, we can write Equation~\eqref{eq:oriprob} as:
\begin{equation}
\begin{aligned}
\label{eq:oriprob_proof}
   \max_{\bm{X}_t} \quad   &  
     \frac{1-\lambda}{|\mathcal{U}|}\sum_{u=1}^{|\mathcal{U}| } \frac{Z(\pi_u) - Z(\pi_u^0)}{Z(\pi_u^*) - Z(\pi_u^0)} + 
    \lambda \cdot \left(  1 - \frac{1}{1 + exp\left(-k\left(G(\bm{e}) - \frac{G(\bm{e}^*) + G(\bm{e}^0)}{2}\right)\right)} \right) \\
   \iff   \max_{\bm{X}_t} \quad & 
     \frac{1-\lambda}{|\mathcal{U}|}\sum_{u=1}^{|\mathcal{U}| } \frac{Z(\pi_u) -1+\exp(q_{\pi_u^*})}{q_{\pi_u}^* - 1+\exp(q_{\pi_u^*})} + 
    \lambda \cdot \left(  1 - \frac{1}{1 + exp\left(-k\left(G(\bm{e}) - \frac{ G(\bm{e}^0)}{2}\right)\right)} \right) \\
  \iff  \max_{\bm{X}_t}  \quad &  
     \frac{1-\lambda}{|\mathcal{U}| }\sum_{u=1}^{|\mathcal{U}| }Z'(\pi_u) + 
    \lambda \cdot G'(\bm{e}) \\
    \text{s.t.} \quad
 &\text{Constraint (9-12)},
    \end{aligned}
\end{equation}
where $g_0=G(\bm{e}^0)$ is the unacceptable level of fairness, which can be adjusted based on the fairness requirement.
\end{proof}

\section{Proof of Theorem~\ref{theo:dual}}

\begin{proof}
For the original optimization problem in Problem~\eqref{eq:oriprob}, we move the constraints to the objective using a vector of Lagrange multipliers $\bm{\mu}\in \mathbb{R}^{|\mathcal{P}|}$:

\begin{equation}
    \begin{aligned}
Y_{OPT} = &  \max_{x_t \in \mathcal{X}_t} \min_{ \bm{\mu}\in \mathcal{D}}\left[ \sum_{u} \frac{1-\lambda}{|\mathcal{U}|}Z'(\pi_u)+ \lambda G'(\bm{e}) + \bm{\mu}^{\top}\left(\bm{e}^\top-\sum_{t=1}^{T} \mathbf{A}^{\top} \bm{X}_{t} \bm{p}\right)\right] \\
 \leq Y_{Dual} = & \min _{\bm{\mu} \in \mathcal{D}}\left[\max _{\bm{X}_{t} \in \mathcal{X}_t}\left[\frac{1-\lambda}{|\mathcal{U}|}Z'(\pi_u)  -\bm{\mu}^{\top} \sum_{t=1}^{T} \bm{A}^{\top} \bm{X}_{t} \bm{p}\right]+\max _{\bm{e} \in \Delta^\mathcal{P}}\left(\lambda G'(\bm{e}) + \bm{\mu}^{\top} \bm{e}\right)\right] \\
 = & \min _{\bm{\mu} \in \mathcal{D}} \left[Z^{*}\left(\bm{A} \bm{\mu} \right)\right]+G^{*}(-\bm{\mu}) ,
   \end{aligned}
    \end{equation}
    where $\mathcal{D}=\{\bm{\mu} | W^*(-\bm{\mu}) \leq \infty\}$ is the feasible region of dual variable $\boldsymbol{\mu}$ for which the conjugate of the regularized is bounded. $Z_t^{*}(\cdot)$ and $W^{*}(\cdot)$ are the conjugate functions.
\end{proof}

\section{Additional Experiments}

\subsection{Experiment Setups}
\subsubsection{Evaluation Metrics}
Besides the metrics selected in the experiment in Section~\ref{sec:experiment}, we also utilize additional metrics to evaluate the effectiveness of our methods.
For user individual fairness, we use the variance and MinMax Ratio of user individual accuracy as the metric. The variance of accuracy (Var) measures the degree of deviation in the accuracy of different users, which is defined as follows:
\begin{equation}
\text{Var}@K=\frac{1}{n^2} \sum_{k=1}^n \sum_{l>k}\left(\text{NDCG}_{u_k}@K-\text{NDCG}_{u_l}@K\right)^2.
\end{equation}

The higher the Var@K value, the more unfair the individual user accuracy is.

\begin{figure}  
    \centering
   \centering
        \begin{subfigure}{\linewidth}
	\centering
	\includegraphics[width = \linewidth]{fig/legend.pdf}
	\end{subfigure}
    \begin{subfigure}{\linewidth}
	\centering
	\includegraphics[width = \linewidth]{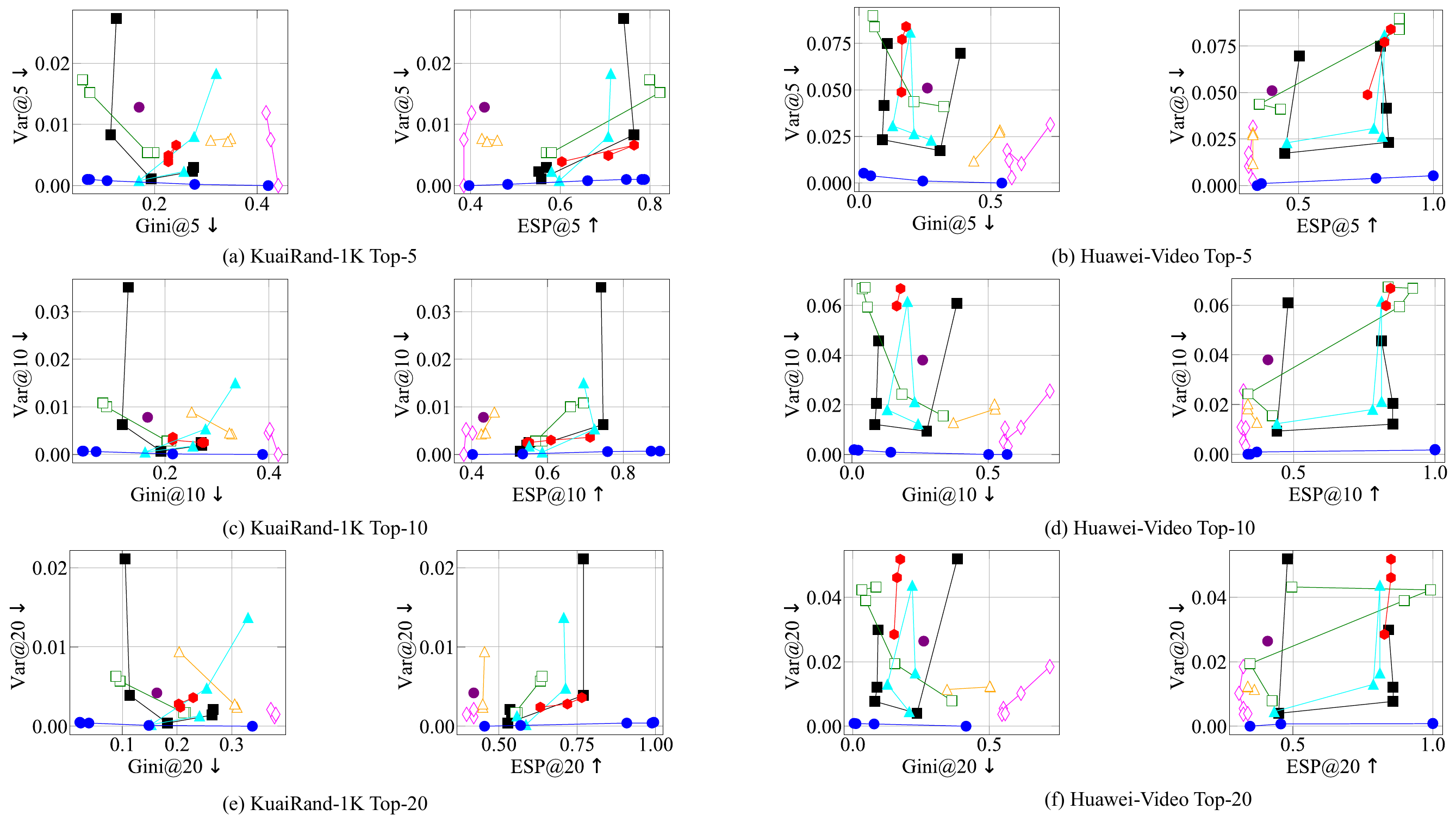}
	\end{subfigure}
    \caption{The Pareto frontierof individual fairness (Var@K) and provider fairness (Gini@K and ESP@K) on two different datasets with top-$10$. X-axis shows Gini@K metric and ESP@K metric, while Y-axis shows Var@K metric. $\uparrow$ means higher values are better and $\downarrow$ favors lower values.}
    \label{fig:pareto_var}
\end{figure}

\subsection{Experimental Results}

We also conduct experiments to verify that our method can ensure both provider fairness and individual fairness on different datasets using Variance to measure individual unfairness. Figures~\ref{fig:pareto_var} illustrate the Pareto frontiers of individual user fairness and provider fairness levels.

From Figure~\ref{fig:pareto_var}, we observe BankFair+, effectively ensures both individual fairness and provider fairness. By adjusting the hyperparameters, our method can achieve Pareto-optimal individual user fairness at different provider fairness levels. As illustrated in Figure~\ref{fig:pareto_var}, we can draw similar conclusions for Var@K. Regardless of the level of provider fairness, the user individual unfairness level of BankFair+ remains extremely low (i.e., Var@K is below 0.002 across different ranking sizes and datasets), while other baselines can only achieve such a level of individual accuracy under bad provider fairness (e.g., Gini@K>0.2 and ESP@K < 0.6). This result demonstrates that our method can significantly ensure provider fairness while maintaining similar recommendation quality for different users, meaning it ensures recommendation accuracy.

\newpage
\bibliographystyle{ACM-Reference-Format}
\balance
\bibliography{ref}
    
\end{document}